\documentclass[twocolumn]{aastex631}

\let\tablenum\relax

\usepackage{amsmath,amssymb,bm}
\usepackage{physics}
\usepackage{mathtools}
\usepackage{microtype}
\usepackage{xcolor}
\begin{document}

\title{The Wave-Regulated Precursor of a Near-Parallel Interplanetary Shock \\ Observed by \textit{Parker Solar Probe}}

\correspondingauthor{I.~C.~Jebaraj}
\email{immanuel.c.jebaraj@gmail.com}

\author[0000-0002-0606-7172]{I.~C.~Jebaraj}
\affil{Department of Physics and Astronomy, University of Turku, FI-20014 Turun yliopisto, Finland}

\author[0000-0001-6016-7548]{L.~Colomban}
\affil{Space Sciences Laboratory, University of California, Berkeley, CA 94720, USA}

\author[0000-0001-6427-1596]{O.~V.~Agapitov}
\affil{Space Sciences Laboratory, University of California, Berkeley, CA 94720, USA}
\affil{Astronomy and Space Physics Department, National Taras Shevchenko University of Kyiv, 03127 Kyiv, Ukraine}

\author[0000-0003-1236-4787]{M.~Gedalin}
\affiliation{Department of Physics, Ben Gurion University of the Negev, Beer-Sheva 84105, Israel}

\author[0000-0001-6360-1987]{M.~A.~Malkov}
\affil{Department of Astronomy and Astrophysics, University of California, San Diego, La Jolla, CA 92093, USA}
\affil{Eureka Scientific, 2452 Delmer St Oakland, CA 94602, USA}

\author[0000-0001-6589-4509]{A.~Kouloumvakos}
\affil{The Johns Hopkins University Applied Physics Laboratory, Laurel, MD 20723, USA}

\author[0000-0002-1349-3663, gname=Edin, sname=Husidic]{E.~Husidic}
\affiliation{Department of Physics and Astronomy, University of Turku, FI-20014 Turun yliopisto, Finland}

\author[0009-0001-4841-1103]{S.~Mondal}
\affiliation{Department of Physics and Astronomy, Queen Mary University of London, London E1 4NS, UK}

\author[0009-0003-9858-5720]{S.~Yadav}
\affil{Department of Physics and Astronomy, University of Turku, FI-20014 Turun yliopisto, Finland}

\author[0000-0001-6344-6956]{N.~Wijsen}
\affil{Centre for mathematical Plasma Astrophysics, KU Leuven, 3001 Leuven, Belgium}

\date{\today}

\begin{abstract}
Diffusive shock acceleration, at shocks from coronal mass ejections to supernova-remnant blast waves, presupposes a scattering wave field that the accelerated particles themselves maintain. This self-regulation has not been resolved in situ. We report Parker Solar Probe observations of a fast ($\simeq$2800~km\,s$^{-1}$), near-parallel interplanetary shock at 0.24~AU on 2023 March 13 and separate its upstream wave field into four families, a classification not made before at a fast shock near the Sun. Right-hand and left-hand circularly polarized families over a common wavenumber band, with a field-aligned linearly polarized family, are cyclotron-resonant with the suprathermal-to-MeV protons streaming from the shock: the beam drives the field that scatters it, and the measured mean free path, half the precursor scale, leaves the beam anisotropic enough to sustain the drive. Outside this loop lies a weak, oblique, linearly polarized component, a few per cent of the wave power, resolved here for the first time at an in situ foreshock. Its in-phase density and field-magnitude fluctuations identify the compressive part as fast magnetosonic and shift the cyclotron-resonance energies of the resonant families by up to 13\% along the precursor. Acceleration at shocks inside 0.3~AU is governed upstream, in a foreshock the shock builds for itself.
\end{abstract}


%------------------------------------------------------------
\section{Introduction}
\label{sec:intro}
%------------------------------------------------------------

Collisionless shocks in space plasmas, from supernova-remnant blast waves to coronal mass ejections (CMEs), are the dominant accelerators of high-energy particles \citep{Kennel85,Krasnoselskikh13,Bykov19book,Gedalinreview26}. In quasi-parallel geometries, where the upstream field is nearly aligned with the shock normal, the shock cannot be separated from its foreshock \citep{Kennel81}: the region upstream that is magnetically connected to the ramp, so accelerated particles stream out, drive instabilities, and are scattered back by the waves those instabilities generate. This coupling sets the diffusion coefficient that governs upstream transport, and with it the efficiency of diffusive shock acceleration \citep{Krymskii77,Axford77,Bell78,Blandford78,Drury83}. How the foreshock is built, and how its internal physics shapes the structure seen at the ramp, remain observationally unresolved.

\begin{figure}[!t]
  \centering
  \includegraphics[width=0.5\textwidth]{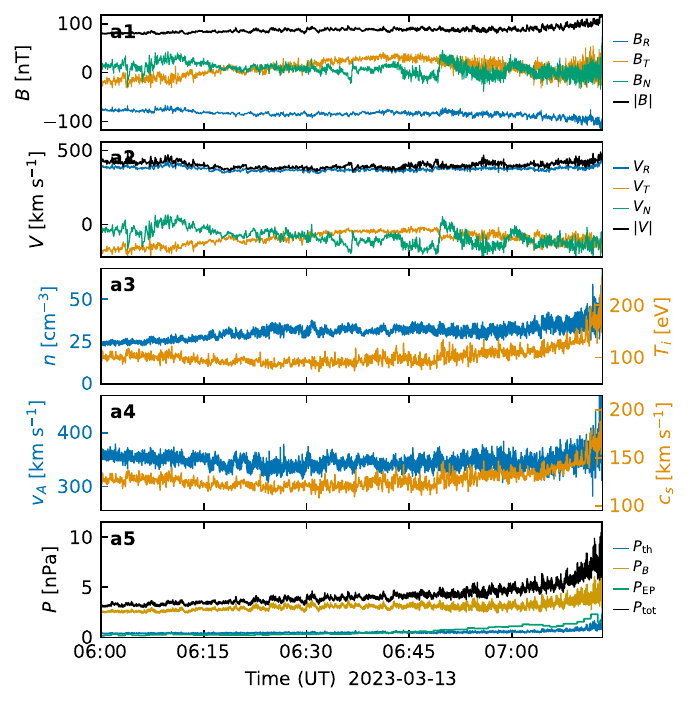}
  \caption{Upstream plasma overview over 06:00--07:13~UT on 2023 March 13; the interval
  ends at the shock crossing at 07:13:13~UT.
  \textit{(a1)} Radial-Tangential-Normal (RTN) magnetic-field components $B_R,B_T,B_N$
  (blue, orange, green) and total $|\bm{B}|$ (black).
  \textit{(a2)} RTN solar-wind velocity $V_R,V_T,V_N$ and total $|\bm{V}|$ (colors as in a1).
  \textit{(a3)} Electron density $n$ (blue, left axis) and ion temperature $T_i$
  (orange, right axis).
  \textit{(a4)} Alfv\'{e}n speed $v_A$ (blue, left axis) and ion sound speed $c_s$
  (orange, right axis).
  \textit{(a5)} Thermal pressure $P_{\rm th}$ (blue), magnetic pressure $P_B$
  (gold), energetic-particle pressure $P_{\rm EP}$ from IS$\odot$IS/EPI-Lo (green),
  and total $P_{\rm tot}=P_{\rm th}+P_B+P_{\rm EP}$ (black).}
  \label{fig:overview}
\end{figure}

Empirical knowledge of the quasi-parallel foreshock comes almost entirely from Earth's bow shock \citep{Eastwood05}. There, beam-driven ultra-low-frequency waves \citep{HoppeRussell83} can steepen into shocklets and short large-amplitude magnetic structures \citep[SLAMS;][]{HadaKennelTerasawa87,Schwartz91,Burgess05, Raptis26}. In strongly driven cases these structures can take over the shock front, making the transition a three-dimensional patchwork of nonlinear features, the front continually reformed by the structures arriving at it \citep{Lembege04,Caprioli14}. The first in situ test of quasi-linear foreshock theory \citep{Lee83} at an interplanetary (IP) shock was carried out by \citet{Kennel86} with the International Sun-Earth Explorer 3 (ISEE-3) at 1~AU. Recent observations of an extreme CME-driven shock close to the Sun revealed a coherent shock transition with no patchwork decomposition and a local rotation of the strongly amplified ramp field toward a quasi-perpendicular configuration \citep{Jebaraj24}, consistent with two-fluid theory \citep{Gedalin98}. Coherent transitions are therefore an alternative endpoint of quasi-parallel geometry, and which endpoint a shock reaches is set upstream, by how far its foreshock fluctuations steepen.

The shock-accelerated ions are predicted to sustain the foreshock through coupled resonant and compressive processes. The field-aligned beam escaping the ramp \citep{Malkov98b} excites cyclotron-resonant electromagnetic waves through the ion/ion instability \citep{Gary85,Vainio03}, and those waves scatter the beam \citep{Lee83,Lagage83,Malkov01}. This resonant scattering sets the parallel mean free path $\lambda_\parallel$ and the upstream diffusion coefficient $\kappa(E)$ that control diffusive shock acceleration \citep{KulsrudPearce69,Skilling75,Vainio14}. The same diffusion builds an exponentially decaying upstream pressure profile. Its gradient, the cosmic-ray acoustic instability it can drive \citep{DorfiDrury85}, and the ponderomotive stress of the accompanying Alfv\'{e}nic wave field \citep{McKenzieVolk82} are each predicted to compress the thermal plasma. These compressions are too long to resonate with the streaming ions, so the plasma responds to them as a fluid \citep{Axford77,Drury81,DruryFalle86} and they modulate the local Alfv\'{e}n speed and proton cyclotron frequency. That modulation feeds back on the resonant scattering and on the cosmic-ray-modified ramp \citep{Haggerty20,Caprioli20}.

At an IP shock the resonant and compressive responses remain unresolved in detail. The quasi-linear coupling has since been tested against wave power spectra \citep{Bamert04}; the resonant families themselves have not been resolved in polarization and wavenumber. The non-resonant compressive response to the energetic-particle gradient, the steady compression of \citet{Drury81} and the acoustic instability it can drive \citep{DorfiDrury85,DruryFalle86}, has never been resolved as a wave field in the foreshock of an IP shock: fast-mode fluctuations have been identified upstream of one laminar subcritical crossing~\citep{LuttrellRichter87}, a regime without a backstreaming beam, the steepened compressive structures of supercritical foreshocks are beam-driven \citep{Raptis26}, and at 1~AU the small amplitude of the predicted response leaves it indistinguishable from ambient solar-wind turbulence \citep{Borovsky19,Zhao21}. Parker Solar Probe \citep[PSP;][]{Fox2016} has now crossed tens of inner-heliospheric shocks at sub-AU distances \citep{Kruparova25}, and both wave fields are within reach at PSP/FIELDS time resolution. We use the IP shock crossed by PSP on 2023 March 13, with supporting measurements from PSP/SWEAP \citep{Kasper16} and IS$\odot$IS \citep{McComas16} (instruments detailed in Appendix~\ref{app:instr}), to resolve the upstream wave field of its foreshock and the energetic-particle precursor that drives it. The field separates into four wave families, a classification not made before at a fast IP shock near the Sun. The resonant electromagnetic families and their organization across the suprathermal-to-MeV beam supply the scattering that underpins diffusive shock acceleration; the beam drives the same families that scatter it, so the scattering field grows with the precursor and the foreshock develops as a consequence of the acceleration itself. The exception is the non-resonant compressive component, resolved here for the first time at an in situ foreshock; the shock does not drive it, yet it modulates the resonance energies of the families the shock does. The energetic-particle population that builds this precursor, and the maximum rigidity it reaches, are analyzed in a companion paper \citep{Kouloumvakos26}.

\begin{figure*}[!t]
  \centering
  \includegraphics[width=0.9\textwidth]{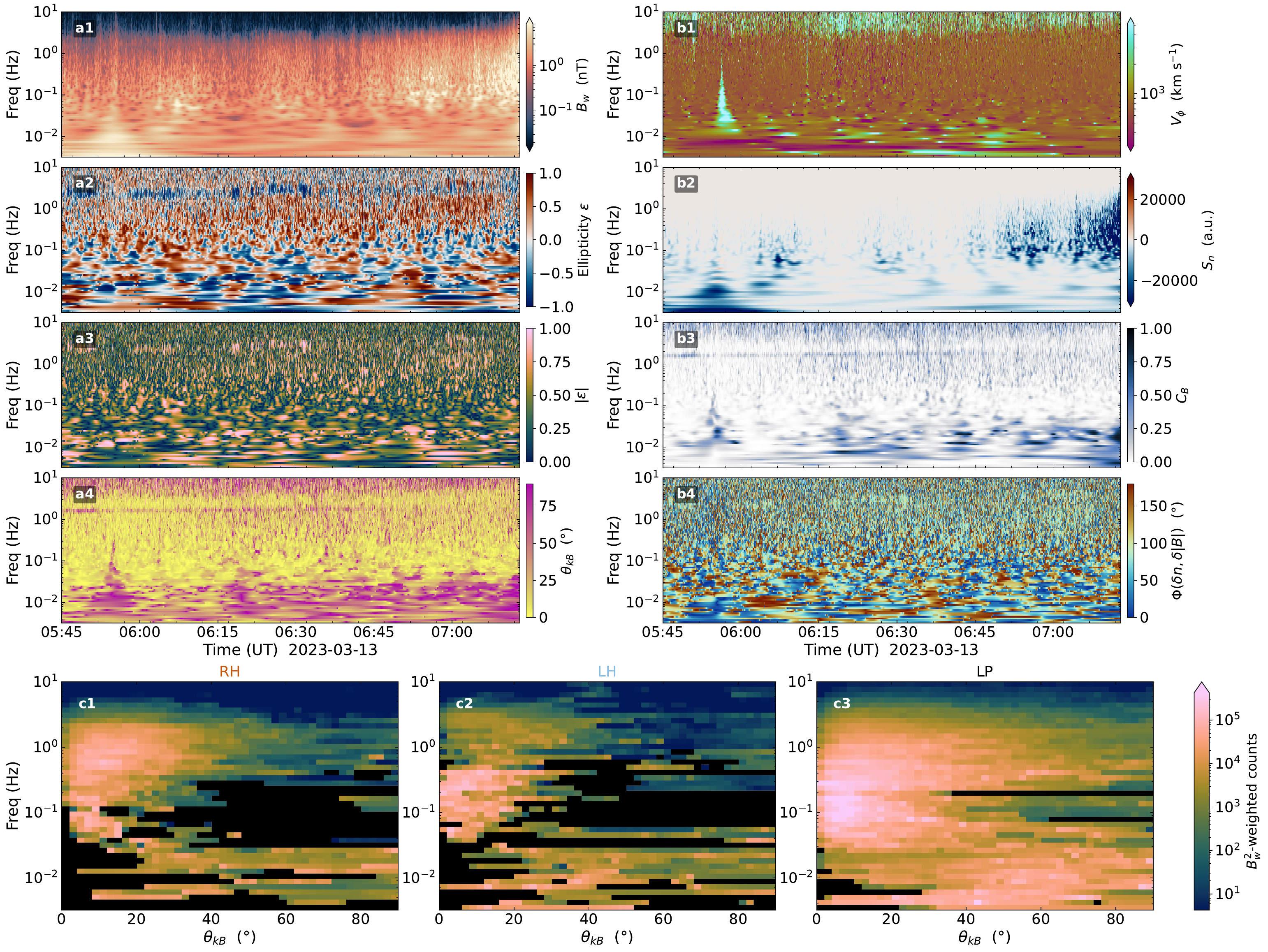}
  \caption{Wave diagnostics for the upstream interval 05:45--07:13~UT on 2023 March 13.
  Panels a1--a4 (left column) and b1--b4 (right column) are time--frequency
  spectrograms over the full interval; panels c1--c3 (bottom row) are
  $B_w^2$-weighted occurrence maps in $(\theta_{kB},f)$ integrated over the inner
  foreshock (06:15--07:13~UT).
  \textit{(a1)} Wavelet magnetic wave amplitude $B_w$ (nT).
  \textit{(a2)} Signed ellipticity $\epsilon$ (positive for right-hand, negative for
  left-hand).
  \textit{(a3)} $|\epsilon|$.
  \textit{(a4)} Propagation angle $\theta_{kB}$ between the minimum-variance
  direction and the local mean field (deg).
  \textit{(b1)} Spacecraft-frame phase velocity $V_\varphi=2\pi f/k$
  (km\,s$^{-1}$).
  \textit{(b2)} Signed shock-normal Poynting flux $S_n$ (a.u.).
  \textit{(b3)} Magnetic compressibility $C_B$.
  \textit{(b4)} Density--field cross-phase $\Phi(\delta n,\delta|\bm{B}|)$ (deg).
  \textit{(c1)} RH occurrence map ($\epsilon\geq +0.7$).
  \textit{(c2)} LH occurrence map ($\epsilon\leq -0.7$).
  \textit{(c3)} LP occurrence map ($|\epsilon|\leq 0.3$).}
  \label{fig:waveoverview}
\end{figure*}

%------------------------------------------------------------
\section{A fully developed foreshock}
\label{sec:overview}
%------------------------------------------------------------

The 2023 March 13 event, observed at $\sim$0.24~AU, has shock-frame upstream speed $u_1\simeq 2400\ \mathrm{km\,s^{-1}}$, obliquity $\theta_{Bn}\simeq 8^\circ$, and Alfv\'enic Mach number $M_A=u_1/v_A\simeq 7.5$ for the representative upstream $v_A\simeq 320\ \mathrm{km\,s^{-1}}$; the measured $v_A$ in the precursor body is $340$--$360$~km\,s$^{-1}$ (Figure~\ref{fig:overview}a4), for which $M_A\simeq 7$. PSP crosses the ramp at 07:13:13~UT, and the first upstream wave activity appears at $\sim$06:00~UT. The inner foreshock, where the wave field is fully developed, starts at $\sim$06:15~UT and extends to the ramp, spanning $\sim$58~min, or $\sim$12~$R_\odot$ ($\sim$0.06~AU) at the measured upstream speed. Fully developed means stationary: the family statistics and the pressure gradient hold steady across the interval, and the fluctuations stay linear throughout (Section~\ref{sec:generation}). All family statistics and derived quantities are evaluated over this inner interval, where $u_1$ is steady, so that time corresponds to shock-normal distance through $x=u_1(t_{\rm shock}-t)$; the spectrograms of Figure~\ref{fig:waveoverview} start at 05:45~UT to include the outer foreshock. This near-parallel geometry places the upstream mean field $\bm{B}_0$ nearly along the shock normal $\hat{\bm{n}}$, so propagation angles $\theta_{kB}$ measured against $\bm{B}_0$ and against $\hat{\bm{n}}$ are equivalent throughout.

Panels a1--a5 of Figure~\ref{fig:overview} characterize the upstream plasma and field. The mean field is quasi-radial throughout (a1) with $B_R<0$ (toward-sector polarity, $\bm{B}_0$ pointing sunward), and $|\bm{B}|\simeq80$~nT through the outer foreshock, approaching $90$~nT by 06:30 and reaching $\simeq100$~nT only in the final $\sim$5~min. The proton cyclotron frequency therefore varies from $f_{ci}\simeq1.2$~Hz in the outer foreshock to $\simeq1.5$~Hz close to the ramp. The bulk solar-wind flow is steady and predominantly radial, with $V_{\rm sw}\simeq 390$~km\,s$^{-1}$ (a2). The Alfv\'{e}n speed $v_A\simeq 320$--$360$~km\,s$^{-1}$ exceeds the sound speed $c_s\simeq 120$--$140$~km\,s$^{-1}$ (a4), with $c_s=(\gamma k_B(T_i+T_e)/m_p)^{1/2}$ and $\gamma=5/3$. The curve in a4 is evaluated from the measured $T_i\simeq 90$--$100$~eV alone, because $T_e\lesssim 40$~eV is unresolved over this interval; including $T_e$ would raise it by $10$--$20\%$, which the representative $c_s\simeq 140$~km\,s$^{-1}$ adopted below allows for. The electron density and ion temperature in a3 yield an upstream number density $n\simeq 30\ \mathrm{cm^{-3}}$ and a plasma beta $\beta\equiv 2\mu_0 nk_B T_i/B_0^2\simeq 0.2$. With the representative $c_s\simeq 140$~km\,s$^{-1}$ and $v_A\simeq 320$~km\,s$^{-1}$ adopted throughout, the fast magnetosonic speed is $c_f=(c_s^2+v_A^2)^{1/2}\simeq 350$~km\,s$^{-1}$; the measured mid-interval pair ($c_s\simeq 125$, $v_A\simeq 345$~km\,s$^{-1}$) gives the same $c_f$ within $5\%$. The sonic and fast magnetosonic Mach numbers are then $M_s\equiv u_1/c_s\simeq 17$ and $M_f\equiv u_1/c_f\simeq 6.9$. The measured $v_A$ varies by only a few per cent across the precursor, so $M_A\simeq 7$--$7.5$ throughout and the shock stays firmly super-fast \citep{Jebaraj24,Jebaraj24b,Wijsen25,Dresing25}.

The energetic-particle pressure $P_{\rm EP}$ from IS$\odot$IS/EPI-Lo grows by close to an order of magnitude across the inner foreshock (Figure~\ref{fig:overview}a5). The thermal ($\propto nT$) and magnetic ($\propto|\bm{B}|^2$) pressures rise by less than a factor of two over the same interval, so $P_{\rm EP}$ has the largest dynamic range in the upstream budget.
\begin{figure*}[!t]
  \centering
  \includegraphics[width=0.85\textwidth]{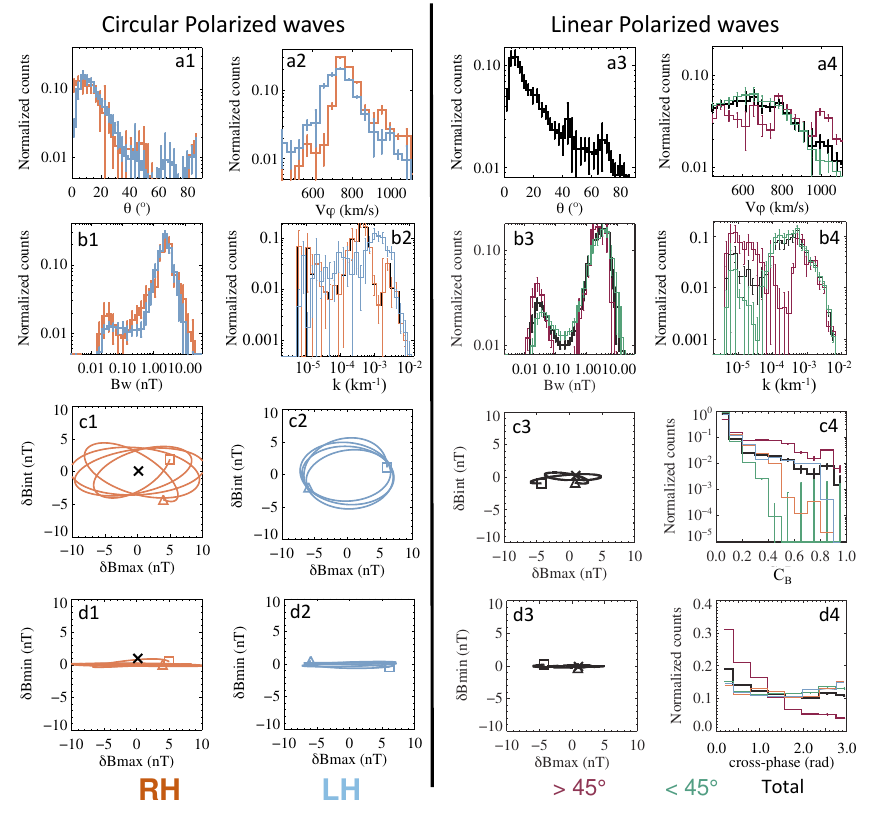}
  \caption{Statistics of the four upstream wave families, integrated over 06:15--07:13~UT.
  Columns 1 and 2 show the circularly polarized families, RH (orange) and LH
  (blue), overlaid in the histogram rows (a, b) and separated in the hodogram
  rows (c, d), RH in column 1 and LH in column 2. Columns 3 and 4 show the
  linearly polarized class, split at $\theta_{kB}=45^\circ$ into LP-OB (maroon)
  and LP-FA (green), with the combined distribution in black (legend: LP (all)).
  Classes are defined by the ellipticity partition at $|\epsilon|=0.2$.
  \textit{(a1, a3)} Propagation angle $\theta_{kB}$ (deg) for the circular families
  (a1) and for the linear population (a3).
  \textit{(a2, a4)} Spacecraft-frame phase velocity $V_\varphi$ (km\,s$^{-1}$) for
  the circular families (a2) and for the linear population (a4).
  \textit{(b1, b3)} Wave amplitude $B_w$ (nT) for the circular families (b1) and for
  the linear population (b3).
  \textit{(b2, b4)} Wavenumber $k$ (km$^{-1}$) for the
  circular families (b2) and for the linear population (b4).
  \textit{(c1--c3)} Representative hodograms in the
  $(\delta B_{\rm max},\delta B_{\rm int})$ plane for RH (c1), LH (c2),
  and the combined linear population (c3), with $\delta B_{\rm max}$,
  $\delta B_{\rm int}$, $\delta B_{\rm min}$ the maximum-, intermediate-, and
  minimum-variance components of the fluctuation. In each hodogram the cross marks the mean
  field and the square and triangle mark the two endpoints of the trace.
  \textit{(c4)} Magnetic-compressibility $C_B$ distribution for the two linear families, with the circular families overplotted.
  \textit{(d1--d3)} Hodograms in the $(\delta B_{\rm max},\delta B_{\rm min})$ plane
  for RH (d1), LH (d2), and the combined linear population (d3).
  \textit{(d4)} Density--field cross-phase $\Phi(\delta n,\delta|\bm{B}|)$ (rad) for
  the two linear families, with the circular families overplotted.}
  \label{fig:stats}
\end{figure*}

%------------------------------------------------------------
\section{Identification of four wave families}
\label{sec:families}
%------------------------------------------------------------

%For each time--frequency bin, a local magnetic spectral matrix is constructed by averaging the wavelet cross-products $\tilde{B}_i\tilde{B}_j^\ast$ over a sliding window spanning six wave periods. Wave properties are then derived from this locally averaged spectral matrix.

We decompose the upstream wave-field fluctuations using a continuous Morlet wavelet transform evaluated at the native magnetic-field cadence ($\sim$293 Hz). The data are described in Appendix~\ref{app:instr} and the methods in Appendix~\ref{app:methods}. We then evaluate the wave amplitude $B_w$, the signed ellipticity $\epsilon$ and its magnitude $|\epsilon|$, the propagation angle $\theta_{kB}$, the spacecraft-frame phase velocity $V_{\varphi}$, the signed shock-normal Poynting flux $S_n$, the magnetic compressibility $C_B$, and the density--field cross-phase $\Phi(\delta n,\delta|\bm{B}|)$. The corresponding time-frequency spectrograms occupy panels a1--a4 and b1--b4 of Figure~\ref{fig:waveoverview}. 

The signed ellipticity assigns each time--frequency bin to a polarization class. The occurrence maps of Figure~\ref{fig:waveoverview}(c1--c3) use the strict cuts $\epsilon\geq+0.7$ for right-hand (RH), $\epsilon\leq-0.7$ for left-hand (LH), and $|\epsilon|\leq0.3$ for linearly polarized (LP), leaving the transitional bands unclassified so that the maps show only the most clearly polarized waves. The family statistics of Figure~\ref{fig:stats}, the dispersion histograms of Figure~\ref{fig:dispersion}, and the power fractions given below all use a complete partition at $|\epsilon|=0.2$, which leaves no bin unassigned.
Because the estimator returns only the resultant polarization in each bin, it cannot by itself distinguish a directly excited linear mode from a coherent superposition of co-located right- and left-hand circular packets at the same $(k,V_\varphi)$. The LP class is split by propagation angle alone, at $\theta_{kB}=45^\circ$, into two families, an oblique one above that angle and a field-aligned one below it. Obliquity is not itself a measure of compression, so the density--field cross-phase is held out of the split and used as an independent test of it (Figure~\ref{fig:stats}d4).
The spacecraft-frame wave energy flux is dominantly anti-sunward across the resolved band, away from the shock and along the backstreaming beam that drives the waves through the cyclotron-resonant ion/ion instability of Section~\ref{sec:generation}. This anti-sunward flux gives $S_n<0$ in Figure~\ref{fig:waveoverview}b2 (sign convention in Appendix~\ref{app:methods}). Its sign does not by itself fix the plasma-frame propagation sense, because the solar wind advects the packets faster than they propagate ($V_{\rm sw}>v_A$), so the wave energy is carried anti-sunward in the spacecraft frame for either sense.

The propagation sense follows instead from the Doppler-shift geometry. The measured phase speed obeys $V_\varphi=V_\varphi^{\rm pl}+\bm{V}_{\rm sw}\cdot\hat{\bm{k}}$, so the plasma-frame speed and the projection of the solar-wind velocity onto the wave vector add. For the quasi-parallel transverse families $\hat{\bm{k}}$ lies close to the radial direction and $V_\varphi^{\rm pl}\simeq v_A$, so anti-sunward propagation predicts $V_{\rm sw}+v_A\simeq 710$~km\,s$^{-1}$ and sunward propagation $V_{\rm sw}-v_A\simeq 70$~km\,s$^{-1}$. The measured $V_\varphi\simeq 700$--$800$~km\,s$^{-1}$ selects the anti-sunward branch, and the kinetic-family wave vectors therefore lie parallel to the beam velocity and to $\bm{V}_{\rm sw}$. For the oblique population the projection $\bm{V}_{\rm sw}\cdot\hat{\bm{k}}$ is smaller and lowers the expected $V_\varphi$; the measured spread on that population is dominated by the single-spacecraft estimator uncertainty and does not test this (Appendix~\ref{app:methods}). Because $\omega^{\rm pl}$ and $\bm{k}\cdot\bm{V}_{\rm sw}$ add for anti-sunward propagation, the spacecraft-frame rotation sense equals the plasma-frame sense, and the measured handedness is not Doppler-reversed.
The shock advances at $V_{\rm shock}=u_1+V_{\rm sw}\simeq 2790$~km\,s$^{-1}$, so the super-Alfv\'{e}nic upstream flow ($M_A\gtrsim 7$) advects the packets back into the shock within the precursor.

The $B_w^2$-weighted occurrence maps in $(\theta_{kB},f)$ space (Figure~\ref{fig:waveoverview}c1--c3) separate the three polarization classes. The circular populations occupy systematically lower propagation angles than the linear class, which extends to larger angles and to lower frequencies. Within the LP class the propagation angle separates a quasi-parallel population that dominates both the occurrence and the power, peaked near $10^\circ$ (Figure~\ref{fig:stats}a3), from a minor oblique component beyond $45^\circ$, and the two sides of that boundary differ in density--field cross-phase and in dispersion (Figures~\ref{fig:stats} and \ref{fig:dispersion}). The upstream wave field therefore decomposes into four families, each named by the polarization and propagation angle that define it: right-hand circular (RH; $\sim$36\% of the $B_w^2$-weighted power over the inner foreshock), left-hand circular (LH; $\sim$37\%), and the two linearly polarized families, oblique (LP-OB; $\sim$6\%) and field-aligned (LP-FA; $\sim$21\%). The names label observed populations, and the four occupy only two plasma branches: the fast magnetosonic branch, whose parallel limit is the R-mode, and the Alfv\'{e}n branch, whose parallel limit is the L-mode \citep{Stix1992}. Which branch each family occupies follows below from the hodograms, the dispersion, and the resonance that drives it \citep{Gary85,Wilson09,Wilson12,Oka19}, and is collected in Table~\ref{tab:families}. The families differ in how they couple to the ions. The RH, LH, and LP-FA families are resonant, coupling to individual ions through the cyclotron condition $k_\parallel v_\parallel\simeq\Omega_{ci}$, with $k_\parallel$ the field-aligned wavenumber and $\Omega_{ci}$ the proton cyclotron frequency, which ties their wavelength to the resonant ion. For the suprathermal-to-MeV beam this gives $2\pi v_\parallel/\Omega_{ci}\sim3\times10^3$--$2\times10^4$~km, well above the proton inertial length $d_i=c/\omega_{pi}\simeq40$~km, with $\omega_{pi}$ the proton plasma frequency. The LP-OB family is non-resonant, its wavelength $\lambda\sim10^5$~km placing the cyclotron resonance on ions far more energetic than the beam, so the waves couple to the bulk plasma as a fluid.

The RH and LH families occupy a common kinematic region and differ in their sense of circular polarization and their wave-power peak position. Both are transverse, low-compressibility ($C_B\lesssim 0.2$) populations peaked below $15^\circ$, with roughly half the occurrence below $20^\circ$ and a tail to large angles (Figure~\ref{fig:stats}a1), and with spacecraft-frame frequencies in the range 0.05--1~Hz. Their wavenumbers
\begin{equation*}
      k_{\rm RH,LH}\simeq 3\times 10^{-4}\text{--}10^{-2}\ \mathrm{km^{-1}},
\end{equation*}
peak in occurrence near $10^{-3}$~km$^{-1}$ for the LH family and at a few$\times 10^{-4}$~km$^{-1}$ for the RH family (b2), while the wave power of both families is centered near $10^{-3}$~km$^{-1}$ (Figure~\ref{fig:dispersion}); these scales set the resonant energies derived in Section~\ref{sec:generation}. The spacecraft-frame phase velocities peak at 700--800~km\,s$^{-1}$ (full range 500--1100~km\,s$^{-1}$), statistically indistinguishable between the two families (a2), clustered at the Doppler-shifted Alfv\'enic asymptote $v_A+V_{\rm sw}$ where the solar-wind advection dominates the single-spacecraft phase speed. The amplitude distribution peaks at $B_w\simeq 1$--$3.5$~nT in both ($\delta B/B_0\sim 1$--$4.5\%$), with a weak second peak (RH) or low shelf (LH) at $B_w\simeq 0.02$--$0.05$~nT, at the floor of the analysis (Section~\ref{sec:generation}).

In the $(\delta B_{\rm max},\delta B_{\rm int})$ plane the RH and LH families trace closed ellipses (c1, c2), and the flat $(\delta B_{\rm max},\delta B_{\rm min})$ traces (d1, d2) confirm negligible minimum-variance amplitude, validating the polarization classification. The signed ellipticity, referenced to $\bm{B}_0$ (Appendix~\ref{app:methods}), identifies them with the R- and L-modes, the parallel limits of the fast magnetosonic and Alfv\'{e}n branches.

The linearly polarized population differs from the circular families across the diagnostics. Its hodograms trace a near-flat line in both the $(\delta B_{\rm max},\delta B_{\rm int})$ and $(\delta B_{\rm max},\delta B_{\rm min})$ planes (Figure~\ref{fig:stats}c3, d3). The amplitude distribution (b3) peaks broadly at $B_w\simeq 1$--$6$~nT, over the same analysis floor at $B_w\simeq 0.02$--$0.06$~nT. The propagation-angle distribution (a3) is dominated by the quasi-parallel population near $10^\circ$, with a minor oblique component near $45^\circ$--$55^\circ$ well below it in occurrence and a weak high-obliquity tail. The two families separate in compressibility and density--field cross-phase, while the single-spacecraft phase speed does not resolve them. The LP-FA waves peak at $V_\varphi\simeq 650$--$700$~km\,s$^{-1}$ (a4), at the Doppler-shifted Alfv\'{e}n speed within the bin width, where the projection $\bm{V}_{\rm sw}\cdot\hat{\bm{k}}$ is nearly constant and $V_\varphi$ is well determined. On the LP-OB population the single-spacecraft electric-to-magnetic estimate of $V_\varphi$ loses accuracy: the unresolved $E_R$ and $E_\parallel$ components enter for oblique $\bm{k}$, and the weak transverse magnetic component inflates the estimate (Appendix~\ref{app:methods}). The intrinsic plasma-frame separation $c_f-v_A=c_s^2/(c_f+v_A)\simeq 30$~km\,s$^{-1}$ (the perpendicular limit; $\simeq23$~km\,s$^{-1}$ at the LP-OB median obliquity) is too small to set the families apart by phase speed. The magnetic compressibility is enhanced in the oblique family, its distribution extending to $C_B\simeq 0.6$ while the quasi-parallel one stays at $C_B\lesssim 0.1$ (c4). A single coplanar mode at the LP-OB median obliquity would have $C_B=\sin^2\theta_{kB}\simeq0.75$, so the family mixes a compressive component with transverse power at the same angles; the compressive part is the one isolated by the density--field cross-phase (Appendix~\ref{app:scales}). In the $k$--$f$ dispersion of Figure~\ref{fig:dispersion} (columns~3 and~4) the LP-OB family occupies the long-wavelength ($k\sim 3\times10^{-5}$~km$^{-1}$) fast magnetosonic band, while the LP-FA family tracks the Doppler-shifted $v_A$ curve.

The LP-OB family is compressive. Its density--field cross-phase $\Phi(\delta n,\delta|\bm{B}|)$ concentrates near zero (Figure~\ref{fig:stats}d4), with $42\%$ of the LP-OB bins of the complete partition within $22.5^\circ$ of in-phase ($48\%$ power-weighted) against $3\%$ for the circular families, the signature of fast magnetosonic compression. An antiphase minority, a quarter of the LP-OB bins, does not survive the in-phase projection of Appendix~\ref{app:scales}. The wavenumbers
\begin{equation*}
  k_{\rm LP\text{-}OB}\simeq 1\times 10^{-5}\text{--}3\times 10^{-4}\ \mathrm{km^{-1}},
\end{equation*}
peak near $3\times10^{-5}$~km$^{-1}$ (b4). Its amplitude distribution peaks at $B_w\simeq 1$--$6$~nT ($\delta B/B_0\sim 1$--$7.5\%$).

A broad occurrence peak at $k\sim 10^{-4}$--$10^{-3}$~km$^{-1}$ (Figure~\ref{fig:stats}b4), overlapping both the LH and RH peaks, places the LP-FA family at the wavenumbers of the circular families. Its $\delta n$--$\delta|\bm{B}|$ cross-phase is incoherent (d4), and its amplitude distribution peaks where the LP-OB family does (b3). Table~\ref{tab:families} summarizes the measured properties of the four families.

\begin{deluxetable*}{lcccc}
% \tabletypesize{\footnotesize}
\tabletypesize{\small}
\tablecaption{Measured properties of the four upstream wave families. Frequencies $f$ and phase velocities $V_\varphi$ are in the spacecraft frame, and $\delta B/B_0$ is reported for the dominant amplitude peak. LP-OB and LP-FA are the two families of the linearly polarized class, separated at $\theta_{kB}=45^\circ$. Wavenumber peaks are occurrence statistics and the power-weighted structure is shown in Figure~\ref{fig:dispersion}; the wave-branch and coupling-regime rows are the identifications established in Sections~\ref{sec:families} and \ref{sec:generation}; the LP-OB branch entry applies to its in-phase compressive part. \label{tab:families}}
\tablewidth{0pt}
\tablehead{
  \colhead{Property} &
  \colhead{RH} &
  \colhead{LH} &
  \colhead{LP-OB} &
  \colhead{LP-FA}
}
\startdata
Polarization sense                 & Right-hand                    & Left-hand                     & Linear                              & Linear                              \\
Coupling regime                    & Resonant (ion-kinetic)        & Resonant (ion-kinetic)        & Non-resonant (fluid)                & Resonant (ion-kinetic)                         \\
$\theta_{kB}$ (peak)               & $\lesssim 15^\circ$           & $\lesssim 15^\circ$           & $45^\circ$--$55^\circ$              & $10^\circ$--$15^\circ$              \\
$\theta_{kB}$ (median)             & $\simeq 21^\circ$             & $\simeq 23^\circ$             & $\simeq 61^\circ$                   & $\simeq 16^\circ$                   \\
$f$ (Hz)                           & $0.05$--$1$                   & $0.05$--$1$                    & $0.003$--$0.03$                     & $0.01$--$1$                         \\
$k$ (peak, km$^{-1}$)   & few$\times 10^{-4}$           & $\sim 10^{-3}$                & $\sim 3\times 10^{-5}$              & $10^{-4}$--$10^{-3}$ (broad)        \\
$k$ (range, km$^{-1}$)        & $3\times 10^{-4}$--$10^{-2}$   & $3\times 10^{-4}$--$10^{-2}$   & $1\times 10^{-5}$--$3\times 10^{-4}$ & $10^{-4}$--$10^{-2}$                \\
$V_\varphi$ (km\,s$^{-1}$)         & $700$--$800$                  & $700$--$800$                  & $\sim 600$--$1100$                  & peak $\simeq 650$--$700$            \\
$B_w$ (floor, nT)         & $0.02$--$0.05$               & $0.02$--$0.05$               & $0.02$--$0.06$                     & $0.02$--$0.06$                     \\
$B_w$ (dominant, nT)               & $1$--$3.5$                   & $1$--$3.5$                   & $1$--$6$                           & $1$--$6$                           \\
$\delta B/B_0$ (dominant)          & $1$--$4.5\%$                  & $1$--$4.5\%$                  & $1$--$7.5\%$                        & $1$--$7.5\%$                        \\
$C_B$                              & $\lesssim 0.2$                & $\lesssim 0.2$                & $\lesssim 0.2$ (to $\sim$0.6)      & $\lesssim 0.1$                      \\
$\delta n$--$\delta|\bm{B}|$ phase & incoherent                    & incoherent                    & $\approx 0$                         & incoherent                          \\
Wave branch                        & Fast magnetosonic (R-mode) & Alfv\'{e}n (L-mode)   & Fast magnetosonic             & Alfv\'{e}n (shear limit)      \\
\enddata
\end{deluxetable*}

%------------------------------------------------------------
\section{Excitation of foreshock fluctuations}
\label{sec:generation}
%------------------------------------------------------------

% The cyclotron-resonant ion/ion instability drives the RH, LH, and LP-FA families, and a non-resonant compressive process drives the LP-OB family.

\subsection*{Resonant kinetic excitation}

Suprathermal ions streaming upstream from the shock have a field-aligned velocity distribution that resonates with electromagnetic waves through the cyclotron condition, written in the plasma frame,
\begin{equation}
  \omega - k_\parallel v_\parallel = \pm\Omega_{ci},
  \label{eq:resonance}
\end{equation}
where $\Omega_{ci}=eB_0/m_p$ is the proton cyclotron angular frequency. The $-\Omega_{ci}$ resonance drives the R-mode (right-hand polarized, magnetosonic--whistler branch), and the $+\Omega_{ci}$ resonance drives the L-mode (left-hand polarized, Alfv\'{e}n--ion-cyclotron branch) \citep[for the broader family of beam-driven electromagnetic ion/ion instabilities see][]{Gary91}. The free energy is the super-Alfv\'{e}nic field-aligned drift of the resonant population ($v_\parallel \gg v_A$), with growth drawing on the excess of the beam drift over the wave phase speed at the resonant velocity \citep{Gary91}. For a cold, strictly field-aligned beam, linear theory gives the strongest growth on the R-mode, through the right-hand resonant ion/ion instability. The L-mode resonance selects ions moving opposite to the beam, so comparable L-mode growth requires the backward hemisphere of the distribution to be populated, the hot-beam condition of \citet{Gary85}. The partially isotropized precursor supplies both hemispheres (Section~\ref{sec:discussion}). At Earth's bow shock both helicities occur, with intrinsic left-hand events the exception \citep{HoppeRussell83}; here the two branches carry near-equal power. Solving Equation~\eqref{eq:resonance} for the resonant wavenumber gives
\begin{equation}
  k_{\rm res} \simeq \frac{\Omega_{ci}}{|v_\parallel \mp v_A|},
  \label{eq:kres}
\end{equation}
where the upper sign refers to the R-mode resonance with ions streaming along the wave and the lower sign to the L-mode resonance with ions streaming against it. For $v_\parallel\gg v_A$, $k_{\rm res}\simeq\Omega_{ci}/v_\parallel$. With $B_0\simeq 80$~nT and parallel velocities $v_\parallel\sim 4\times 10^{3}$--$3\times 10^{4}$~km\,s$^{-1}$ spanning 0.1--5~MeV, Equation~\eqref{eq:kres} yields $k_{\rm res}\sim 2\times 10^{-4}$--$2\times 10^{-3}$~km$^{-1}$. The RH occurrence peak at a few$\times 10^{-4}$~km$^{-1}$ corresponds to cyclotron resonance with outward-streaming protons of 1--3~MeV, and the LH occurrence peak near $10^{-3}$~km$^{-1}$ to sunward-moving protons near 300~keV, both abundant in the backstreaming EPI-Lo population. The power-weighted median wavenumbers are $1.0\times10^{-3}$~km$^{-1}$ (LH) and $1.4\times10^{-3}$~km$^{-1}$ (RH) (Figure~\ref{fig:dispersion}), in resonance with protons of 160--310~keV. The growth rate follows the resonant streaming flux, which falls steeply with energy, so the power concentrates at the high-wavenumber end of the band and remains weak at the longer wavelengths.

The linear growth rate of the cyclotron-resonant ion/ion instability further restricts the unstable spectrum to quasi-parallel propagation. The resonance condition $k_\parallel v_\parallel = \pm \Omega_{ci}$ couples wave growth to the parallel-beam free energy through $k_\parallel = k\cos\theta_{kB}$. At oblique angles the wave polarization becomes a mixture of compressive and transverse components, and the cyclotron coupling is reduced. Linear growth is strongest at parallel propagation and falls with obliquity, persisting tens of degrees off-axis for a hot beam \citep{Gary85,Gary91}; the broad pitch-angle distribution of the backstreaming population feeds several parallel resonances at once, and convective transit of a wave packet through the precursor preferentially amplifies the most-unstable parallel modes, acting as an angular filter. The observed RH and LH occurrence peaks below $15^\circ$ (Figure~\ref{fig:waveoverview}c1, c2), the LP-FA peak at $10^\circ$--$15^\circ$, and the absence of a distinct oblique population in the kinetic-family bands are the angular signatures of this combined growth and selection.

The dominant $B_w$ peak of the RH and LH families (Figure~\ref{fig:stats}b1) strengthens toward the ramp (Figure~\ref{fig:waveoverview}a1) in step with the rising $P_{\rm EP}$ (Figure~\ref{fig:overview}a5), tracking the e-foldings a packet accrues over the precursor transit. The low-amplitude structure in the same distributions is confined to $4$--$9$~Hz at $B_w\simeq0.02$--$0.05$~nT ($\delta B/B_0\sim0.05\%$), the same floor recorded in all four families (Table~\ref{tab:families}); it is the analysis noise floor at high frequency, not a second wave population.

The LP-FA family sits at the Doppler-shifted Alfv\'{e}n speed, its $k$--$f$ distribution tracking that curve (Figure~\ref{fig:dispersion}, column~4) and its spacecraft-frame phase speed clustering at $V_\varphi\simeq 700$~km\,s$^{-1}$ (Figure~\ref{fig:stats}a4). Neither diagnostic separates the two branches here, because both asymptote to that speed on the low-frequency wing, where a coherent superposition of the dispersion-displaced RH and LH packets registers as a linearly polarized signal. The branch assignment rests instead on the drive: the family is a resonant shear-Alfv\'{e}n response to the streaming suprathermal beam \citep{Vainio03}, of the same gyroresonant kind as the RH and LH waves, with a contribution from ambient solar-wind Alfv\'{e}nic fluctuations advected inward by the upstream flow. Because its broad wavenumber distribution overlaps both the RH and LH peaks at amplitudes comparable to either, the family augments cyclotron-resonant pitch-angle scattering across the $\sim$300~keV to few-MeV range that dominates $P_{\rm EP}$.

\subsection*{The compressive LP-OB family}

The compressive part of the LP-OB family lies on the fast magnetosonic branch. Its density and field-magnitude fluctuations are in phase (cross-phase $\Phi\simeq0$; Figure~\ref{fig:stats}d4), which excludes the competing slow ion-acoustic branch, whose density perturbation is in antiphase and which is strongly Landau-damped because the measured $T_i\sim90$~eV exceeds the electron temperature. The cyclotron resonance at its peak wavelength falls on relativistic protons with negligible flux (Appendix~\ref{app:transport}), outside the band the beam instabilities grow, and its obliquity suppresses what growth remains.

A steady energetic-particle pressure gradient convected at the super-fast speed ($M_f\simeq6.9$) forces only a weak compression,
\begin{equation}
\frac{\delta n}{n}\sim\frac{P_{\rm EP}}{\rho_0\,(u_1^2-c_f^2)}=\frac{\mathcal{N}}{M_f^2-1},\qquad \mathcal{N}=\frac{P_{\rm EP}}{\rho_0 c_f^2},
\label{eq:Nmax}
\end{equation}
where ram inertia exceeds magnetosonic stiffness by $M_f^2$. Across the inner foreshock $\mathcal{N}$ rises from $\simeq0.04$ to $\simeq0.3$, so this response spans $0.1$--$0.6\%$ and stays an order of magnitude below the observed amplitude at every distance. The steady cosmic-ray-gradient response of \citet{Drury81}, generalized to the magnetosonic stiffness in Appendix~\ref{app:forced}, therefore cannot supply it. At the $M_A\simeq4$ of the simulations of \citet{Giacalone26} the response would rise to $0.3$--$2.4\%$; the exclusion stands on the measured $u_1=2400\pm300$~km\,s$^{-1}$ \citep{Jebaraj24}. The resonant Alfv\'enic waves compress the plasma through the gradient of their magnetic pressure \citep{Hollweg71}, which averaged over the carrier is $P_A=\langle b_A^2\rangle/2\mu_0$, with $b_A$ the Alfv\'{e}nic wave amplitude. The envelope of that pressure is steady in the shock frame, so the plasma sweeps through it at $u_1$ and the response is ram-limited in the same way (Appendix~\ref{app:forced}),
\begin{equation}
\frac{\delta n}{n}\simeq \frac12\,\frac{v_A^2}{u_1^2-c_f^2}\left(\frac{b_A}{B_0}\right)^2,
\label{eq:pond}
\end{equation}
which is $\simeq0.006\%$ where $b_A/B_0$ reaches $\simeq8\%$ in the strongest resonant packets near the ramp, more than two orders of magnitude below the LP-OB amplitude. Envelope variations on the packet scale travel at $v_A$ relative to the plasma and force a stronger response, $\tfrac12(b_A/B_0)^2\,v_A^2/(v_A^2-c_s^2)$, which reaches $0.4\%$ at the strongest packets near the ramp and stays more than an order of magnitude below the observed amplitude at every distance (Appendix~\ref{app:forced}). The energetic-particle pressure gradient drives compressive instabilities \citep{ZankAxfordMcKenzie90,DruryFalle86,Capanema26}, but a steady gradient grows them by at most $\simeq0.05$ e-foldings over the transit (Appendix~\ref{app:forced}). Neither the resonant nor the non-resonant ion/ion beam instability produces it \citep{Gary85,Gary91}. The cyclotron resonance of the field-aligned beam grows quasi-parallel, predominantly transverse waves and builds the RH, LH, and LP-FA families; its growth falls with obliquity \citep{Gary91}, and the convective filter that sets their occurrence peaks below $\theta_{kB}\simeq15^\circ$ suppresses what remains at the LP-OB obliquities. The non-resonant branch grows only for drifts above $\sqrt{2}$ times the beam thermal speed or for dense beams \citep{Gary85}, and the hot, tenuous precursor population meets neither condition. The near-equal LH power is itself the signature of that hot beam. Parametric decay of the transverse families places its compressive daughters at and above the pump wavenumber, well separated from the LP-OB band, and its ion-acoustic daughter is heavily Landau-damped at the measured $T_i>T_e$ (Appendix~\ref{app:forced}). Anisotropy-driven compressions of the mirror type anticorrelate $\delta n$ and $\delta|\bm{B}|$ and are excluded by the in-phase relation of Figure~\ref{fig:stats}d4.

\begin{figure}[!t]
  \centering
  \includegraphics[width=\columnwidth]{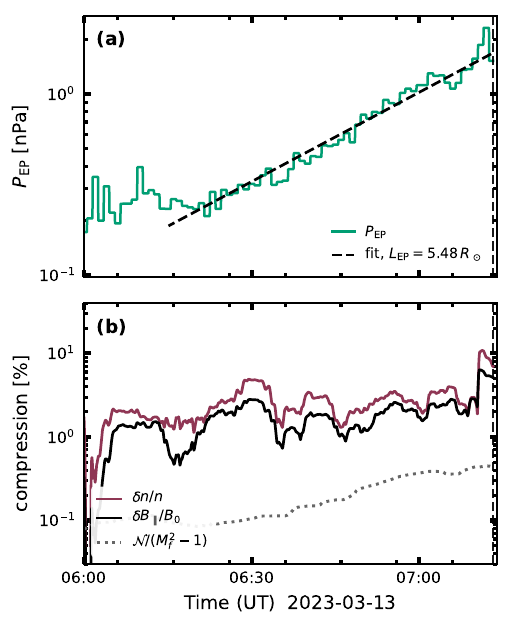}
  \caption{The energetic-particle precursor and the compressive amplitude over
  06:00--07:13~UT on 2023 March 13; the time axis corresponds to shock-normal distance
  through $x=u_1(t_{\rm shock}-t)$, and the dashed line marks the shock.
  \textit{(a)} Energetic-particle pressure $P_{\rm EP}$ (green) with its exponential fit
  of scale length $L_{\rm EP}\simeq5.48\,R_\odot$ (black, dashed).
  \textit{(b)} The compressive amplitudes $\delta n/n$ (maroon) and
  $\delta B_{\parallel}/B_0$ (black), the in-phase coherent components in the LP-OB band (Appendix~\ref{app:scales}),
  with the energetic-particle-gradient forced response $\mathcal{N}/(M_f^2-1)$ (gray, dotted).}
  \label{fig:pressures}
\end{figure}

With the local channels excluded, the remaining possibility is ambient fast-mode turbulence advected from the solar wind. Its amplitude is close to constant through the body of the precursor and rises only in the last few minutes before the ramp (Figure~\ref{fig:pressures}b), so it does not follow the order-of-magnitude growth of $P_{\rm EP}$. The single-spacecraft phase speed is too poorly determined on this oblique population (Appendix~\ref{app:methods}) to separate a freely propagating mode from a convected one. A second spacecraft would settle this, and the origin of the rise in the final minutes, through the spatial growth of the amplitude.

The observed $\delta n/n$ and $\delta B_\parallel/B_0$, band-passed to the LP-OB frequencies and projected onto the component in phase with the density, are set against the precursor pressure in Figure~\ref{fig:pressures}. The projection removes the transverse families, which lie mainly at higher frequencies and are incoherent with $\delta n$ (Appendix~\ref{app:scales}). The absence of SLAMS-like jumps in the magnetic field (Figure~\ref{fig:overview}a1) places the present system below the nonlinear regime. The wave field grows toward the ramp, the resonant families in step with $P_{\rm EP}$ and the compressive $\delta B_\parallel/B_0$ from $1.7\%$ through the body to $6.3\%$ in the final minutes, and reaches the shock without saturating. \citet{Raptis26} resolve the near-shock region of an IP foreshock, within $\simeq150\,d_i$ of the ramp, where compressive structures initiate but do not evolve into mature SLAMS. The present measurement covers the extended precursor instead, $2\times10^{5}\,d_i$ at $M_A\simeq7$ and $\theta_{Bn}\simeq8^\circ$, and finds the wave field linear across all of it.

%------------------------------------------------------------

\section{The foreshock as a coupled system}
\label{sec:discussion}
%------------------------------------------------------------

The four wave families resolved here couple to the energetic particles in both directions. The accelerated particles drive the resonant families, and all four act back on the particles. The transverse RH, LH, and LP-FA families scatter the suprathermal precursor through cyclotron resonance, the resonant energy at parallel wavenumber $k_\parallel$ being
\begin{equation}
  E_{\rm res} \simeq \frac{1}{2}m_p\!\left(\frac{\Omega_{ci}}{k_\parallel}\right)^{2}.
  \label{eq:Eres}
\end{equation}
This non-relativistic form holds for $k_\parallel\gtrsim 2\times 10^{-4}$~km$^{-1}$, and across the kinetic-family band ($k_\parallel\simeq 3\times10^{-4}$--$10^{-2}$~km$^{-1}$) it spans a few keV to a few MeV, covering the backstreaming population that dominates $P_{\rm EP}$.

The scattering these families provide diffuses the precursor; the parallel mean free path follows the quasi-linear estimate \citep{Lee83,Vainio00},
\begin{equation}
  \lambda_\parallel \sim r_g\!\left(\frac{B_0}{\delta B}\right)^{2},
  \label{eq:qlambda}
\end{equation}
with $r_g=p/(eB_0)$. The amplitude entering Equation~\eqref{eq:qlambda} is the resonant one, the transverse power within an e-fold of wavenumber about the resonance: $\delta B_{\rm res}/B_0\simeq 5$--$6\%$ at the 1--2.4~MeV resonances (Appendix~\ref{app:methods}), giving $\lambda_\parallel\simeq 0.7$--$1.7\,R_\odot$; the total transverse amplitude, $10\%$, would lower this by a factor of three. The precursor scale $L_{\rm EP}\simeq 5.48\,R_\odot$ returns an empirical $\kappa_{\rm eff}=u_1 L_{\rm EP}\simeq 9.2\times 10^{15}$~m$^{2}$\,s$^{-1}$, an average over the energies that dominate the energetic-particle pressure rather than a single channel, and implies $\lambda_\parallel=3\kappa_{\rm eff}/v\simeq 1.9$--$2.9\,R_\odot$ over the same energies. The estimate and the measurement agree in magnitude; their energy scalings differ (Appendix~\ref{app:transport}). Two to three scattering lengths then span the precursor, so the beam isotropizes only partially and its residual field-aligned drift keeps feeding the waves. In a unidirectional wave field a single helicity resonates in only one hemisphere of pitch angle; scattering the full distribution requires both helicities over a common wavenumber band, and both are present here. This scattering field underpins the diffusive acceleration and the maximum rigidity the precursor reaches, analyzed for this event in the companion paper \citep{Kouloumvakos26}.

The compressive LP-OB family couples to the same particles non-resonantly. Its oblique propagation and long peak wavelength place its own cyclotron resonance on relativistic protons ($\gtrsim$100~MeV; Appendix~\ref{app:transport}), so it does not scatter the suprathermal population that dominates $P_{\rm EP}$, but its in-phase compressive modulation of the field magnitude, the $\delta B_\parallel/B_0\simeq 1.7\%$ that rises to $\simeq 6.3\%$ toward the ramp (Figure~\ref{fig:pressures}), shifts the local $v_A$ and $\Omega_{ci}$ and, through Equation~\eqref{eq:Eres}, the resonant energies of the kinetic families. Since $E_{\rm res}\propto(B/k_\parallel)^2$ for $\omega\ll\Omega_{ci}$, a field modulation moves them at fixed $k_\parallel$ by $\delta E_{\rm res}/E_{\rm res}\simeq 2\,\delta B_\parallel/B_0$, or $3$--$13\%$ here.

The same compressions transfer energy. Transit-time damping \citep{Schlickeiser98} works through the $n=0$ Landau resonance at the plasma-frame parallel phase speed $c_f(\theta_{kB})/\cos\theta_{kB}\simeq 700$~km\,s$^{-1}$ at the LP-OB median obliquity, spanning $500$--$1500$~km\,s$^{-1}$ across the $50^\circ$--$76^\circ$ spread, too fast for the thermal protons ($v_{{\rm th},i}\simeq 135$~km\,s$^{-1}$) but resonant with the suprathermal ions and, within the electron thermal spread ($v_{{\rm th},e}\simeq 2700$--$3800$~km\,s$^{-1}$ for $T_e=20$--$40$~eV), the electron bulk. Magnetic pumping \citep{Malkov26} heats through the broadband modulation of the magnetic moment once the cyclotron-resonant scattering breaks the cycle-to-cycle reversibility, so each compression leaves the perpendicular motion of the scattered suprathermal population hotter and its temperature rises secularly. Both channels scale as the square of the compression amplitude, so the high-amplitude tail of the LP-OB distribution (Figure~\ref{fig:stats}b3, $\delta B$ up to $\simeq 6$~nT) dominates the energy transfer. At the amplitudes measured here the pumping rate is small against the convective rate $u_1/L_{\rm EP}\simeq 6\times 10^{-4}$~s$^{-1}$, and its square-law dependence makes it competitive in stronger or closer-in shocks.

The energetic-particle pressure builds the precursor, the escaping beams drive the kinetic waves, and those waves scatter the beams back, so the resonant upstream regulates itself with no external prescription of the wave amplitude. The compressive family stands outside that loop, driven by none of the local channels examined in Section~\ref{sec:generation} and Appendix~\ref{app:forced}, and acts on it by moving the resonance energies and by heating. The body force the precursor exerts is small, $a_{\rm EP}=(\mathcal{N}/M_f^2)\,u_1^2/L_{\rm EP}\simeq 5$~m\,s$^{-2}$ at the middle of the precursor against the convective $u_1^2/L_{\rm EP}\simeq 1.5\times 10^3$~m\,s$^{-2}$, so it structures the upstream without altering the bulk shock parameters, and with $P_{\rm EP}/\rho_0u_1^2=\mathcal{N}/M_f^2\simeq 3\times 10^{-3}$ the shock stays weakly cosmic-ray-modified.

The present measurement resolves the resonant field in polarization and wavenumber and adds the compressive component that the ISEE-3 analyses of \citet{Kennel84,Kennel86} recorded only as the power in the field magnitude. \citet{Giacalone26}, modeling the same crossing, independently trace the energetic particles to shock-heated solar wind rather than to pre-existing suprathermal populations. Their field-matched simulations return $M_A\simeq4$ against the $7$--$7.5$ that the in-situ plasma and shock timing give here.
The measured $\mathcal{N}=0.04$--$0.3$ places this shock in the weakly modified regime. As stronger or longer-driven shocks approach $\mathcal{N}$ of order unity, $\delta B/B_0$ across the resonant band grows toward unity and the mean free path collapses toward the gyroradius, raising the energy reachable within the shock residence time. There the nonlinear steepening of the acoustic precursor and non-resonant streaming \citep{Bell04,Malkov06,CaprioliSpit14c} take the system into the high-rigidity regime invoked for galactic cosmic rays.

Within $\sim 0.3$~AU the foreshock spans a substantial fraction of the heliocentric distance, so its energetic-particle pressure, its resonant waves and compressive component, and the heating they drive evolve with the shock that builds them. Acceleration at CME-driven shocks this close to the Sun is therefore governed upstream, in a precursor that models must follow as it develops. It is there that the diffusion coefficient, the partition of energy among species, and the suprathermal seed for injection are set.

\section*{acknowledgments}
I.C.J. and E.H. were supported by the Research Council of Finland (X-Scale, grant No.~371569).
I.C.J. was also supported by FWO grant No.~1295826N. 
I.C.J. acknowledges support from ISSI's ``Visiting Scientist Program''.
M.M. was supported by NASA ATP 80NSSC24K0774 (UCSD), 80NSSC26K0593 (Eureka Sci.), and Fermi 80NSSC25K7346 grants.
A.K. acknowledges financial support from the NASA NNN06AA01C contract, from NASA's LWS grant 80NSSC25K0130, and the NSF SHINE grant 2401162.
%
% \end{acknowledgments}

%------------------------------------------------------------
\appendix
%------------------------------------------------------------

\section{Instrumentation}
\label{app:instr}

The observations are from Parker Solar Probe \citep{Fox2016} during the inbound segment of the 2023 March 13 encounter at $\sim$0.24~AU heliocentric distance.

\subsection{Electromagnetic fields (FIELDS)}
\label{app:instr:fields}

The FIELDS suite \citep{Bale16} provides the magnetic-field, electric-field, and spacecraft-potential measurements. The vector magnetic field $\bm{B}$ is from a redundant pair of three-axis DC fluxgate magnetometers (MAG), returning the three RTN components at up to 293~vectors~s$^{-1}$ during the encounter window. The electric-field instrument consists of four whip antennas extended in the spacecraft heat-shield plane, measuring the two components of $\bm{E}$ transverse to the Sun-pointing axis. The third component, along the Sun-pointing axis and close to radial, is not directly measured; in this quasi-radial event that axis lies close to $\bm{B}_0$. The electron density $n$ is obtained from the FIELDS spacecraft floating potential (Appendix~\ref{app:instr:density}).

\subsection{Solar Wind Electrons, Alphas, and Protons (SWEAP)}
\label{app:instr:sweap}

The SWEAP suite \citep{Kasper16} provides the plasma moments. The Solar Probe ANalyzer for Ions \citep[SPAN-Ai;][]{Livi22} is a top-hat electrostatic analyser with a time-of-flight section that resolves the three-dimensional proton velocity distribution over 20~eV--20~keV at $\sim$3.5~s cadence, from which the ion temperature $T_i$ and the proton bulk velocity $\bm{V}_{\rm sw}$ are obtained. The companion electron analyser SPAN-E \citep{Whittlesey20} resolves the electron velocity distribution and provides the electron temperature $T_e$.

\subsection{Integrated Science Investigation of the Sun (IS$\odot$IS)}
\label{app:instr:isois}

The IS$\odot$IS suite \citep{McComas16} resolves the energetic-particle populations. The Energetic Particle Instrument Low-energy detector \citep[EPI-Lo;][]{Hill17} is a time-of-flight mass spectrometer with 80 separate apertures arranged to give a $\sim$2$\pi$-steradian field of view, returning the proton differential intensities $j(E,\Omega,t)$ from $\sim$20~keV to $\sim$10~MeV across multiple energy and pitch-angle bins at $\sim$10~s cadence.

\subsection{Calibration of the spacecraft-potential electron density}
\label{app:instr:density}
The high-cadence electron density used throughout is derived from the FIELDS spacecraft floating potential $V_{\rm sc}$, which resolves density fluctuations across the MHD and kinetic scales that the particle instruments do not reach. Solar illumination charges the spacecraft positive through photoemission, so the balance of photoelectric and thermal-electron currents makes $V_{\rm sc}$ track the local electron density $n$ through $n\propto e^{-V_{\rm sc}/V_{\rm pe}+C_{\rm b}}$ \citep{Pedersen_1995,Chen_2012a,Mozer_2022,Mondal_2026}, with $V_{\rm pe}$ the photoelectron thermal energy. Here $V_{\rm sc}=-(V_1+V_2+V_3+V_4)/4$ is the mean of the four antenna potentials, the offset $C_{\rm b}$ absorbs the antenna bias, and the ion and secondary-electron currents are negligible. The relation is calibrated against the quasi-thermal-noise density from the Radio Frequency Spectrometer \citep{Pulupa2017}. The potential adjusts to density changes with a time constant $\tau_c\approx C V_{\rm pe}/I_{\rm th,e}$ \citep{Chen_2013b}, with $C$ the spacecraft capacitance and $I_{\rm th,e}$ the thermal-electron current. The method fails above $\tau_c^{-1}\gtrsim 10$~kHz near the Sun, well above the wave frequencies analyzed here ($\lesssim 2$~Hz).

\section{Data analysis}
\label{app:methods}

\begin{figure*}[!t]
  \centering
  \includegraphics[width=0.99\textwidth]{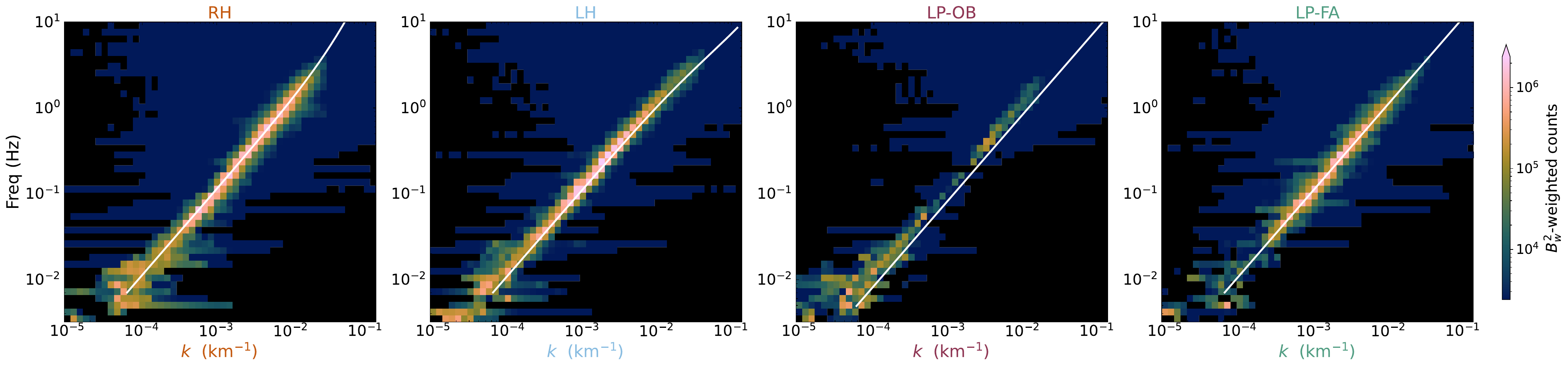}
  \caption{Wavenumber content of the four upstream wave families, RH, LH, LP-OB, and
  LP-FA (columns 1--4). Each panel is a $B_w^2$-weighted two-dimensional
  histogram of spacecraft-frame frequency $f$ against the wavenumber
  $k=2\pi f/V_\varphi$. White curves are the plasma-frame branches (R-mode,
  L-mode, fast magnetosonic at the LP-OB median $\theta_{kB}$, and shear
  Alfv\'{e}n), Doppler-shifted into the spacecraft frame.}
  \label{fig:dispersion}
\end{figure*}

\paragraph{Wavelet decomposition.}
The three RTN components of $\bm{B}$ from MAG are decomposed using a continuous complex Morlet wavelet transform \citep{TorrenceCompo98} with Morlet parameter $\omega_0=10$ for sharper frequency localization than the conventional $\omega_0=6$. The transform is evaluated at the native MAG cadence, yielding time--frequency representations $\tilde{B}_i(t,f)$ over the frequency range $3.2\times10^{-3}$--$10$~Hz. The wave amplitude $B_w(t,f)$ is the magnetic fluctuation amplitude of the time--frequency cell, in nT. The field-magnitude coefficients, $\widetilde{|\bm{B}|}(t,f)$, are obtained using the same transform.
Density coefficients $\tilde{n}(t,f)$ are obtained from the calibrated spacecraft-potential density time series with the same transform. The coefficients $\widetilde{|\bm{B}|}(t,f)$ and $\tilde{n}(t,f)$ enter the density--field cross-phase diagnostic. The magnetic compressibility is instead computed from the projection of the local magnetic spectral matrix along the background magnetic-field direction.
All magnetic wave properties discussed below are derived from a time-averaged magnetic spectral matrix (Equation~\eqref{eq:M_S}) and are therefore defined on the magnetic-field $(t,f)$ grid.
The occurrence maps of Figure~\ref{fig:waveoverview}(c1--c3) and the two-dimensional histograms of Figure~\ref{fig:dispersion} are weighted by the wave power $B_w^2$ rather than by bin counts, because $B_w^2$ enters the quasi-linear scattering rate and hence the diffusion coefficient (Equation~\eqref{eq:qlambda}). The one-dimensional family histograms of Figure~\ref{fig:stats}, and the peak and median propagation angles, phase velocities, amplitudes, and wavenumbers quoted from them, are instead occurrence distributions in bin counts, while the family power fractions are $B_w^2$-weighted.

\paragraph{Spectral matrix and polarization classification.}
For each time--frequency bin, we construct a complex Hermitian magnetic spectral matrix from the MAG wavelet coefficients,
\begin{equation}
S_{ij}(t,f)
=
\left\langle
\tilde{B}_i(t,f)\,
\tilde{B}_j^{*}(t,f)
\right\rangle,
\qquad
i,j\in\{R,T,N\},
\label{eq:M_S}
\end{equation}
where the angle brackets denote a moving average over a sliding window of six wave periods ($6/f$) centered on each sample. We then decompose the normalized spectral matrix by singular value decomposition into singular values $\lambda_2\ge\lambda_1\ge\lambda_0\ge0$ and corresponding singular vectors \citep{McPherron72,Means72,Santolik03,Taubenschuss19}. For a single plane wave, $S_{ij}$ is rank-deficient with $\lambda_0\rightarrow0$. The singular vector associated with $\lambda_0$ defines the wave-normal direction $\hat{\bm{k}}(t,f)$, with the $\pm180^\circ$ ambiguity resolved at the band level using the sign of the radial Poynting flux.

The wave-normal angle shown in Figures~\ref{fig:waveoverview} and \ref{fig:stats} is
\begin{equation}
\theta_{kB}(t,f)
=
\arccos\!\bigl(|\hat{\bm{k}}\cdot\hat{\bm{B}}_0|\bigr),
\label{eq:M_thetakb}
\end{equation}
where $\hat{\bm{B}}_0$ is the local mean magnetic-field direction and is obtained by low-pass filtering the magnetic-field time series with a cutoff frequency equal to one-fifth of the analyzed wave frequency. The signed ellipticity $\epsilon(t,f)$ is computed after rotating the spectral matrix into the local wave frame $(\bm{e}_1,\bm{e}_2,\hat{\bm{k}})$, where $\bm{e}_1$ is the projection of $\hat{\bm{B}}_0$ onto the plane perpendicular to $\hat{\bm{k}}$ and $\bm{e}_2=\hat{\bm{k}}\times\bm{e}_1$. Denoting the rotated matrix by $S'$, we take
\begin{equation}
\epsilon(t,f)
=
\frac{2\,\mathrm{Im}\!\left(S'_{12}\right)}{S'_{11}+S'_{22}}\,.
\label{eq:M_eps}
\end{equation}
Linear polarization corresponds to $|\epsilon|=0$, while circular polarization corresponds to $|\epsilon|=1$. Equation~\eqref{eq:M_eps} is signed by the sense of rotation about $\hat{\bm{k}}$, while the R- and L-mode identity is the sense of rotation about $\bm{B}_0$. The ellipticity used throughout is therefore $\epsilon\,\mathrm{sign}(\hat{\bm{k}}\cdot\hat{\bm{B}}_0)$, positive for right-handed and negative for left-handed rotation about $\bm{B}_0$. In this event $\bm{B}_0$ points sunward and $\hat{\bm{k}}$ anti-sunward, so the factor is $-1$ throughout the interval; the referencing was verified against the frame-free rotation sense $(\delta\bm{B}\times\partial_t\delta\bm{B})\cdot\hat{\bm{B}}_0$ computed from the magnetometer time series.

\paragraph{Phase velocity and Poynting flux.}
The spacecraft-frame phase velocity is estimated from the wavelet-domain electric-to-magnetic-field ratio using Faraday's relation for a plane wave,
\begin{equation}
\hat{\bm{k}}\times\tilde{\bm{E}}(t,f)
=
V_\varphi(t,f)\,\tilde{\bm{B}}(t,f)\,,
\label{eq:M_faraday}
\end{equation}
with $V_\varphi=\omega/k$ the phase speed in the spacecraft frame and $\hat{\bm{k}}$ the wave-normal direction obtained from the magnetic spectral matrix. The FIELDS antennas resolve the two electric-field components $\tilde{E}_x$ and $\tilde{E}_y$ in the heat-shield plane, while the third component lies along the Sun-pointing axis, close to $\hat{\bm{R}}$, and is not directly measured. In the right-handed triad $(\hat{\bm{x}},\hat{\bm{y}},\hat{\bm{R}})$, with $\tilde{\bm{B}}$ rotated into the same frame, the $x$ component of Equation~\eqref{eq:M_faraday} reads
\begin{equation}
V_\varphi\,\tilde{B}_x
=
\hat{k}_y\tilde{E}_R-\hat{k}_R\tilde{E}_y\,,
\label{eq:M_faraday_x}
\end{equation}
with analogous expressions for the other components. For the quasi-parallel transverse families $|\hat{k}_R|\simeq 1$ and $\hat{k}_x,\hat{k}_y\ll 1$, and the unresolved $\tilde{E}_R$ is small, so that $V_\varphi\tilde{B}_x\simeq-\hat{k}_R\tilde{E}_y$. The estimator used throughout is therefore
\begin{equation}
V_\varphi(t,f)
=
\left|
\frac{\tilde{E}_y(t,f)\,\tilde{B}_x^{*}(t,f)}{|\tilde{B}_x(t,f)|^2}
\right|,
\label{eq:M_vphi}
\end{equation}
with the analogous estimate based on $\tilde{E}_x\tilde{B}_y^{*}/|\tilde{B}_y|^2$ used as a consistency check. The two give consistent distributions, and the reported values correspond to $|V_\varphi|$ in the spacecraft frame.
For oblique waves the simplification is less accurate, because $\hat{k}_R$ departs from unity and the unresolved $\tilde{E}_R$ contributes to Equation~\eqref{eq:M_faraday_x} through $\hat{k}_y$. For the LP-OB family $V_\varphi$ is an order-of-magnitude diagnostic and is not used as an identification criterion.
This technique is similar to those presented by \citet{chust2021}, \citet{colomban2024}, and \citet{colomban2025}.
Electric-field fluctuations were obtained from the differential antenna voltages using an effective antenna length of $L_{\rm eff}=1.8$ m, such that $\tilde{E}_{x,y}=\tilde{V}_{x,y}/L_{\rm eff}$. This value is consistent with the effective antenna length reported by \citet{Mozer2020} for low-frequency electric-field measurements on PSP. An uncertainty in $L_{\rm eff}$ scales $V_\varphi$ and $k=2\pi f/V_\varphi$ uniformly across all families and leaves their relative ordering unchanged. Throughout, measured frequencies and phase speeds carrying no superscript are in the spacecraft frame, and a superscript ${\rm pl}$ marks the plasma frame. The plasma-frame phase speed $V_\varphi^{\rm pl}$ relates to the measured $V_\varphi$ through the Doppler shift $V_\varphi=V_\varphi^{\rm pl}+\bm{V}_{\rm sw}\cdot\hat{\bm{k}}$, with $\bm{V}_{\rm sw}$ the solar-wind bulk velocity from SPAN-Ai. The scalar fast-magnetosonic speed $c_f=(c_s^2+v_A^2)^{1/2}$ quoted in the main text is the perpendicular ($\theta_{kB}=90^\circ$) limit. The angle-dependent fast speed entering the LP-OB family is $c_f(\theta_{kB})=\{\tfrac12[c_s^2+v_A^2+\sqrt{(c_s^2+v_A^2)^2-4c_s^2 v_A^2\cos^2\theta_{kB}}]\}^{1/2}$ \citep{Stix1992}, which departs from the scalar value by less than the single-spacecraft phase-speed uncertainty over the observed obliquity range.

We compute the Poynting flux in the spacecraft frame from the complex wavelet spectra $\tilde{\bm{E}}$ from FIELDS and $\tilde{\bm{B}}$ from MAG \citep[following][]{Sundkvist12,Karbashewski23}. Only the radial projection of the Poynting flux,
\begin{equation}
  S_R(t,f) = \frac{1}{\mu_0}\,\mathrm{Re}\!\left[\tilde{E}_T(t,f)\,\tilde{B}_N^{*}(t,f)-\tilde{E}_N(t,f)\,\tilde{B}_T^{*}(t,f)\right]\,,
  \label{eq:M_Sz}
\end{equation}
is independent of the unresolved $E_R$ and reliably determined. For the present near-radial shock ($\theta_{Bn}\simeq 8^\circ$) and toward-sector geometry ($\hat{\bm{n}}\parallel\bm{B}_0$, with $\hat{\bm{n}}\approx-\hat{\bm{R}}$), the shock-normal projection $S_n=\bm{S}\cdot\hat{\bm{n}}\approx-S_R$ to within a few per cent. Figure~\ref{fig:waveoverview}b2 displays this $S_n$.

\paragraph{Resonant-band amplitude.}
The amplitude entering the quasi-linear estimate of Section~\ref{sec:discussion} is the transverse power within an e-fold of wavenumber about the cyclotron resonance, $\delta B_{\rm res}/B_0=[f_{\rm res}\,P_\perp(f_{\rm res})]^{1/2}/B_0$, with $P_\perp$ the Welch spectral density of the two transverse components in field-aligned coordinates over the inner foreshock and $f_{\rm res}$ the Doppler-shifted resonant frequency. Evaluated with the representative $B_0=80$~nT, it is $5$--$8\%$ across the 0.3--2.4~MeV resonances ($f_{\rm res}=0.12$--$0.04$~Hz), and $5$--$6\%$ over the 1--2.4~MeV range used in Section~\ref{sec:discussion}, against a total transverse amplitude of $10\%$ over 0.05--1~Hz.

\paragraph{Theoretical dispersion curves.}
The white curves of Figure~\ref{fig:dispersion} are plasma-frame dispersion relations evaluated with the representative foreshock parameters ($B_0\simeq 80$~nT, $n\simeq 30\ \mathrm{cm^{-3}}$, $v_A\simeq 320$~km\,s$^{-1}$, $c_s\simeq 140$~km\,s$^{-1}$, $d_i\simeq 40$~km) and then Doppler-shifted into the spacecraft frame. For parallel propagation the circular branches are the cold-plasma R- and L-modes \citep{Stix1992},
\begin{equation}
n_{\rm R,L}^2
=
1-\frac{\omega_{pe}^2}{\omega(\omega\mp\Omega_{ce})}-\frac{\omega_{pi}^2}{\omega(\omega\pm\Omega_{ci})},
\qquad
V_\varphi^{\rm pl}=\frac{c}{n_{\rm R,L}},\quad k=\frac{n_{\rm R,L}\,\omega}{c},
\label{eq:M_RL}
\end{equation}
with the upper signs for the R-mode (magnetosonic--whistler, passing through $\Omega_{ci}$ without a stop band and steepening toward the whistler regime) and the lower signs for the L-mode (ion-cyclotron, evanescent above $\Omega_{ci}$, so $V_\varphi^{\rm pl}\to0$ as $\omega\to\Omega_{ci}^-$). Here $\omega_{pe}$, $\omega_{pi}$ are the electron and proton plasma frequencies and $\Omega_{ce}$ is the electron cyclotron frequency. The curve overlaid on the LP-FA panel is $\omega^{\rm pl}=k\,v_A$ (shear Alfv\'{e}n, parallel limit) and that on the LP-OB panel is $\omega^{\rm pl}=k\,c_f(\theta_{kB})$ at the median propagation angle of the LP-OB family. Every curve is then transformed with
\begin{equation}
\omega = \omega^{\rm pl}+\bm{k}\cdot\bm{V}_{\rm sw},
\qquad
V_\varphi = V_\varphi^{\rm pl}+V_{\rm sw}\cos\theta_{kR},
\label{eq:M_doppler}
\end{equation}
and drawn parametrically in $k$ as $f(k)$, with $f=\omega/2\pi$ and $\theta_{kR}$ the angle between $\hat{\bm{k}}$ and the radial direction. Because $k$ is constructed from $f$ and $V_\varphi$, every sample of Figure~\ref{fig:dispersion} satisfies $f=kV_\varphi/2\pi$ identically, so each ridge has unit slope and its width measures the spread of $V_\varphi$ alone. Below $0.3$~Hz, where the power-weighted median wavenumber of every family lies, the R- and L-mode curves lie within $2.5\%$ of the shear-Alfv\'{e}n curve and the three are visually coincident. The LP-OB curve is displaced to larger $k$ by the factor $1.33$ that the $\cos\theta_{kR}$ projection removes from its Doppler term, against a branch-speed difference of $c_f-v_A=23$~km\,s$^{-1}$ at that obliquity. The RH, LH, and LP-FA overlays use $\cos\theta_{kR}\simeq1$ (quasi-parallel), while the LP-OB overlay is projected at that family's median $\theta_{kB}$.

\paragraph{Compressibility and density--field cross-phase.}
The magnetic compressibility plotted in Figures~\ref{fig:waveoverview} and \ref{fig:stats} is computed from the local magnetic spectral matrix as
\begin{equation}
  C_B(t,f) =
  \frac{\hat{\bm{B}}_0^{\,T} S(t,f)\,\hat{\bm{B}}_0}{\mathrm{tr}\,S(t,f)}\,,
  \label{eq:M_CB}
\end{equation}
where $\hat{\bm{B}}_0$ is the local background magnetic-field direction. It measures the fraction of the magnetic fluctuation power parallel to the local mean field, normalized by the total magnetic wave power \citep{GarySmith09,TenBarge12}. The density--field cross-phase plotted in Figures~\ref{fig:waveoverview}b4 and \ref{fig:stats}d4 is computed from the wavelet coefficients of the calibrated density (from the FIELDS spacecraft potential) and of the field magnitude (from MAG),
\begin{equation}
  \Phi(\delta n,\delta|\bm{B}|)(t,f)
  =
  \left|
  \arg\!\left[
  \widetilde{|\bm{B}|}(t,f)\,
  \tilde{n}^{*}(t,f)
  \right]
  \right| .
  \label{eq:M_Phi}
\end{equation}
Values near zero correspond to in-phase density and field-magnitude fluctuations, as expected for fast-mode-like compressions, whereas values near $\pi$ correspond to anticorrelated fluctuations, as expected for slow-mode-like compressions \citep{Howes12}.

\paragraph{Energetic-particle pressure and precursor scale.}
We construct the total upstream energetic-particle pressure plotted in Figures~\ref{fig:overview} and \ref{fig:pressures} from the EPI-Lo proton differential intensities $j(E,\Omega,t)$, adapted from \citet{Lario15}, as
\begin{equation}
  P_{\rm EP}(t) = \frac{1}{3}\sum_E\int p(E)\,j(E,\Omega,t)\,d\Omega\,\Delta E\,,
  \label{eq:M_PEP}
\end{equation}
with $p(E)=c^{-1}\sqrt{E^2+2E m_p c^2}$ the proton momentum at kinetic energy $E$, summed over the resolved suprathermal proton energy bins and over all EPI-Lo apertures. Equation~\eqref{eq:M_PEP} reduces to the isotropic form $(4\pi/3)(2m_p)^{1/2}\int E^{1/2}j\,dE$ of \citet{Lario15} in the non-relativistic limit, and the two agree to within $5\%$ across the inner foreshock. Protons above 300~keV carry $99\%$ of $P_{\rm EP}$. The exponential precursor fit $P_{\rm EP}(x)\simeq P_{\rm EP,s}\exp(-x/L_{\rm EP})$, with $P_{\rm EP,s}$ the pressure extrapolated to the shock, to $P_{\rm EP}(t)$ versus shock-normal distance $x=u_1(t_{\rm shock}-t)$ over the inner foreshock yields $L_{\rm EP}\simeq 5.48\,R_\odot$ and the effective upstream diffusion coefficient $\kappa_{\rm eff}=u_1 L_{\rm EP}\simeq 9.2\times 10^{15}$~m$^{2}$\,s$^{-1}$.

\paragraph{Shock-normal direction.}
The shock-normal direction $\hat{\bm{n}}$ used in the shock-frame projection $x=u_1(t_{\rm shock}-t)$ and the propagation-angle diagnostics comes from a magnetic coplanarity inversion of the field averages over the minutes immediately upstream and downstream of the shock crossing at 07:13:13~UT \citep{Colburn66,Abraham72}, cross-checked with minimum-variance analysis of the magnetic field at the ramp \citep{Sonnerup98} within the framework of \citet{Schwartz98}. The obliquity $\theta_{Bn}\simeq 8^\circ$, the angle between $\hat{\bm{n}}$ and the upstream mean field, is the average of the two estimates. The coplanarity sign convention places $\hat{\bm{n}}$ parallel to the upstream $\bm{B}_0$. In this toward-sector event ($B_R<0$) that points sunward, so $S_n<0$ in Equation~\eqref{eq:M_Sz} corresponds to an anti-sunward Poynting flux, away from the shock and into the upstream.

\section{Scale separation of the compressive amplitude}
\label{app:scales}

\begin{figure}[!t]
  \centering
  \includegraphics[width=0.5\columnwidth]{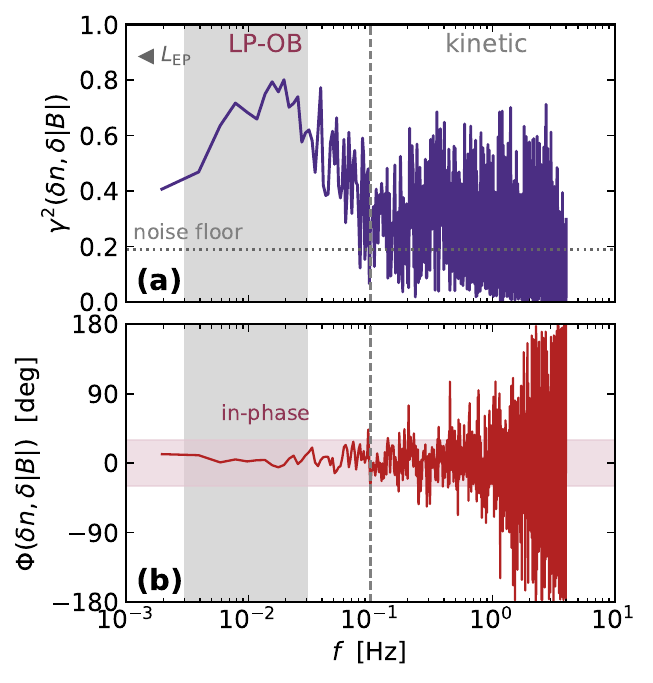}
  \caption{Coherence and cross-phase of density and field magnitude. (a) Density--field-magnitude coherence $\gamma^2(\delta n,\delta|\bm{B}|)$ against spacecraft-frame frequency $f$, formed from gradient-removed fluctuations; the LP-OB band ($0.003$--$0.03$~Hz) is shaded, the dotted line is the noise floor of a phase-randomized control, the dashed line marks the ion-kinetic onset, and the arrow points toward the precursor scale $L_{\rm EP}$, which lies below the plotted frequency range. (b) The signed density--field cross-phase $\Phi(\delta n,\delta|\bm{B}|)$; the shaded band marks $\pm 30^\circ$ about zero.}
  \label{fig:scales}
\end{figure}

The compressive amplitude set against the forced response in Section~\ref{sec:generation} is the fluctuation of a single non-resonant population, separated from the precursor gradient that drives the forced response and from the transverse kinetic families. The separation rests on a hierarchy of frequency scales. The gradient varies over the precursor length $L_{\rm EP}\simeq 5.48\,R_\odot$ (Appendix~\ref{app:methods}), below $f\simeq 10^{-3}$~Hz. The transverse cyclotron-resonant RH and LH families lie at $f\gtrsim 0.05$~Hz (Table~\ref{tab:families}). The LP-FA family extends down to $f\simeq0.01$~Hz, into the upper part of the LP-OB band; it is nearly incompressible ($C_B\lesssim0.1$), its $\delta n$--$\delta|\bm{B}|$ cross-phase is incoherent, and the in-phase projection below removes it. Between the gradient and the kinetic families the compressive LP-OB family occupies $f\simeq 0.003$--$0.03$~Hz, $\lambda\sim 10^5$~km (Figure~\ref{fig:scales}).

The compressive amplitudes in Figure~\ref{fig:pressures} are isolated as follows. The fluctuations $\delta n/n$ and $\delta|\bm{B}|/B_0$ are taken relative to a $900$~s running mean, removing the smooth gradient compression that sets the forced response, and then band-pass filtered to the frequencies between the gradient below and the cyclotron-resonant families above. The fast magnetosonic mode is finally separated from the antiphase, Landau-damped slow branch by retaining only the part in phase with the density. With $a_n$ and $a_B$ the band-passed $\delta n/n$ and $\delta|\bm{B}|/B_0$ and $\langle\cdot\rangle$ a $300$~s running average, the in-phase coherent amplitudes are
\begin{equation}
\frac{\delta n}{n}=\frac{\langle a_n a_B\rangle}{\langle a_B^2\rangle^{1/2}},\qquad
\frac{\delta B_\parallel}{B_0}=\frac{\langle a_n a_B\rangle}{\langle a_n^2\rangle^{1/2}},
\label{eq:fastamp}
\end{equation}
evaluated where the covariance $\langle a_n a_B\rangle>0$. The magnitude fluctuation $\delta|\bm{B}|$ used for $a_B$ equals the compressive parallel component $\delta B_\parallel$ to first order in the wave amplitude, the transverse contribution entering only at second order ($\delta B_\perp^2/2B_0$). Over the precursor the covariance is positive throughout, so the compression is fast at every distance.

The density--field-magnitude coherence $\gamma^2(\delta n,\delta|\bm{B}|)$ peaks at $\simeq 0.8$ near $f\simeq 20$~mHz (Figure~\ref{fig:scales}a), in the LP-OB band, well above the $\simeq 0.2$ noise floor set by the $95$th percentile of a phase-randomized control, and falls through the kinetic range as those families become transverse. The cross-phase (panel b), shown signed rather than as the absolute value of Equation~\eqref{eq:M_Phi}, holds within $\pm 30^\circ$ of zero across the LP-OB band, the fast-mode signature (Table~\ref{tab:families}). Ideal MHD gives $[1-(c_s^2/c_f^2(\theta_{kB}))\cos^2\theta_{kB}]^{-1}$ for the amplitude ratio $|\delta n/n|/|\delta B_\parallel/B_0|$, which is $1.01$--$1.08$ over the $50^\circ$--$76^\circ$ obliquity spread ($1.04$ at the median) and negative on the slow branch. The measured ratio is positive and $\simeq1.4$; its sign places the population on the fast branch, and its offset above the fast-branch value lies within the multiplicative uncertainty of the spacecraft-potential density calibration, which enters $\delta n/n$ directly.

\section{Forced compressive responses}
\label{app:forced}

Section~\ref{sec:generation} compares the driven compressions of the background plasma, each the response of the thermal fluid to a body force evaluated for the representative foreshock state ($\beta\simeq0.2$, $v_A^2/c_s^2\simeq5.2$).

\subsection{Steady energetic-particle gradient}
In the shock frame the thermal plasma flows along the normal $x$ at speed $u$, with the energetic particles acting through the gradient of their scalar pressure $P_{\rm EP}(x)$ \citep{Drury81}; the magnetosonic stiffness below generalizes their gas-dynamic response. Steady mass and momentum balance read $d(\rho u)/dx=0$ and $\rho u\,du/dx+dP_g/dx+dP_{\rm EP}/dx=0$, so $\rho u=\rho_0 u_1$ is conserved. The oblique compressive mode displaces the field as well as the gas, so the thermal stiffness is the fast-magnetosonic one, $dP_g\to\rho c_f^2\,d\ln\rho$ with $c_f^2=c_s^2+v_A^2$. Linearizing about the far-upstream state, $\rho u=\rho_0 u_1$ gives $\delta u/u_1=-\delta\rho/\rho_0$, and the momentum equation reduces to $(u_1^2-c_f^2)\,d(\delta\rho/\rho_0)=dP_{\rm EP}/\rho_0$, hence Equation~\eqref{eq:Nmax},
\begin{equation}
\frac{\delta n}{n}=\frac{P_{\rm EP}}{\rho_0(u_1^2-c_f^2)}=\frac{\mathcal{N}}{M_f^2-1}.
\label{eq:Nmax_app}
\end{equation}
Because $u_1^2\gg c_f^2,c_s^2$, the sound speed of a strictly field-aligned compression gives the same $\delta n/n$ to within $2\%$.

\subsection{Ponderomotive force}
The resonant Alfv\'enic waves oscillate on a fast carrier of wavelength $\lambda\sim 10^3$--$10^4$~km and plasma-frame frequencies well below $f_{ci}$, while their amplitude $b_A$ varies on the slow envelope of the precursor, $L_{\rm EP}\simeq 5.48\,R_\odot$, more than two orders of magnitude longer. Averaging over the carrier leaves the cycle-averaged magnetic pressure $P_A=\langle b_A^2\rangle/2\mu_0$, a function of the envelope coordinate alone, whose gradient is the ponderomotive body force $-\nabla P_A$ on the bulk plasma, the carrier average of the wave magnetic-pressure drive of \citet{Hollweg71}.

This envelope is steady in the shock frame, exactly as $P_{\rm EP}(x)$ is, and the thermal plasma sweeps through it at $u_1$. The forcing speed in the plasma frame is therefore $u_1$. The balance is the one already solved above with $P_{\rm EP}$ replaced by $P_A$, and the response is limited by ram inertia rather than by the sound speed,
\begin{equation}
\frac{\delta n}{n}=\frac{P_A}{\rho_0(u_1^2-c_f^2)}=\frac12\,\frac{v_A^2}{u_1^2-c_f^2}\left(\frac{b_A}{B_0}\right)^2.
\label{eq:pondqs}
\end{equation}
This is Equation~\eqref{eq:pond}. It falls below the quasi-static estimate $P_A/\rho_0 c_s^2$ by the factor $(u_1^2-c_f^2)/c_s^2\simeq 290$, and with $b_A/B_0\simeq 8\%$ toward the ramp it gives $\delta n/n\simeq 6\times 10^{-5}$. The response is a compression in phase with the field magnitude, so the ponderomotive channel is excluded on amplitude rather than on phase.

The precursor-scale envelope is the slowest envelope of the resonant field. The packet envelopes visible in Figure~\ref{fig:waveoverview}(a1) vary on minutes and propagate at $v_A$ relative to the plasma, so the response they force has the denominator $v_A^2-c_s^2$ in place of $u_1^2-c_f^2$,
\begin{equation}
\frac{\delta n}{n}=\frac12\,\frac{v_A^2}{v_A^2-c_s^2}\left(\frac{b_A}{B_0}\right)^2,
\label{eq:pondpk}
\end{equation}
again a compression in phase with the wave pressure because the driver is supersonic ($v_A>c_s$). With $b_A/B_0=3$--$8\%$ this spans $0.06$--$0.40\%$, inside the LP-OB band in frequency and in phase with $\delta|\bm{B}|$. At the strongest packets, near the ramp, the response stays a factor of $\sim$16 below the observed amplitude there; through the body of the precursor it stays more than an order of magnitude below.

\subsection{Cosmic-ray acoustic growth}
The same gradient drives the gas through the two-fluid acoustic instability \citep{DorfiDrury85,DruryFalle86,Capanema26}. Linearizing the gas equations about a static background carrying a constant $\nabla P_{\rm EP}$ gives $\omega^2=c_s^2 k^2+i\,\bm{k}\cdot\nabla P_{\rm EP}/\rho_0$, so a steady gradient suffices. For $k\gg|\nabla P_{\rm EP}|/\rho_0 c_s^2$ the growing root is $\gamma_D=|\nabla P_{\rm EP}|/2\rho_0 c_s$. Over the precursor transit $\tau=L_{\rm EP}/u_1$, with $|\nabla P_{\rm EP}|\simeq P_{\rm EP}/L_{\rm EP}$, this yields $\gamma_D\tau=P_{\rm EP}/2\rho_0 c_s u_1\simeq0.05$ e-foldings at the ramp-end $P_{\rm EP}$, the scale $L_{\rm EP}$ cancelling. The same linearization on the oblique magnetosonic branches \citep{ZankAxfordMcKenzie90,Capanema26} adds the obliquity projection and a polarization factor, and both the fast and the slow root gain less over the transit than the parallel acoustic rate.

\subsection{Parametric decay of the transverse families}
The RH, LH, and LP-FA pumps carry $\delta B/B_0\lesssim4.5\%$ each at $\beta\simeq0.2$. For a parallel pump at $k_0$ the decay daughter carrying the density sits at $k_s=2k_0/(1+c_s/v_A)\simeq1.4\,k_0$, at the pump scale $k_0\sim10^{-3}$~km$^{-1}$ and more than an order of magnitude above the LP-OB occurrence peak, with the antiphase slow-mode polarization that the in-phase projection removes. That daughter is an ion-acoustic wave, heavily Landau-damped at the measured $T_i\simeq90$~eV~$>T_e$, and the pump spans a decade in wavenumber, so the coherent three-wave growth is further reduced by the spread of pump phases. Envelope self-modulation drives the packet-scale response of Equation~\eqref{eq:pondpk} and is bounded by it.

\section{Diffusion--convection transport in the precursor}
\label{app:transport}

The upstream diffusion-convection equation for the energetic-particle distribution $f(x,p)$ in the shock frame \citep{Parker65}, with $x$ increasing into the upstream so that the flow velocity along $x$ is $-u_1$,
\begin{equation}
  -u_1\,\partial_x f = \partial_x(\kappa\partial_x f) + \tfrac{1}{3}(\partial_x u)\,p\,\partial_p f,
  \label{eq:diffconv}
\end{equation}
reduces in the upstream region (where $\partial_x u\simeq 0$ and $\kappa$ varies slowly) to $\kappa\partial_x^2 f+u_1\partial_x f\simeq 0$, whose decaying solution is $f\propto e^{-x/L}$ with $L=\kappa/u_1$. The measured exponential $P_{\rm EP}(x)$ therefore yields a single effective $\kappa_{\rm eff}=u_1 L_{\rm EP}$, set by the MeV energies that dominate $P_{\rm EP}$. Evaluated with this constant $\kappa_{\rm eff}$, the implied parallel mean free path and convective-to-scattering timescale ratio at energy $E$ are
\begin{equation}
  \lambda_\parallel(E) = \frac{3\kappa_{\rm eff}}{v(E)},
  \qquad
  \frac{\tau_{\rm conv}}{\tau_{\rm sc}} = \frac{v(E)^2}{3u_1^2},
  \label{eq:lambda}
\end{equation}
with $\tau_{\rm conv}=L_{\rm EP}/u_1$ the convection time and $\tau_{\rm sc}=\lambda_\parallel/v$ the scattering time. For 1--2.4~MeV protons ($v\simeq 1.4$--$2.1\times 10^4$~km\,s$^{-1}$), $\tau_{\rm conv}/\tau_{\rm sc}\simeq 10$--$27$, placing transport in the diffusive regime. The ratio falls toward unity below $\simeq0.1$~MeV, where the diffusion approximation ceases to hold. The quasi-linear estimate $\lambda_\parallel\sim r_g(B_0/\delta B)^2$ of Section~\ref{sec:discussion}, evaluated with the resonant-band amplitude, agrees with $3\kappa_{\rm eff}/v$ within a factor of four at 1~MeV and within $10\%$ at 2.4~MeV; the two scale oppositely with energy.

The cyclotron-resonant energy at parallel wavenumber $k_\parallel$ and field $B$ is, numerically,
\begin{equation}
  E_{\rm res}\,(\mathrm{MeV}) \simeq 4.79\times 10^{-11}\,\left[\frac{B\,(\mathrm{nT})}{k_\parallel\,(\mathrm{km^{-1}})}\right]^2.
  \label{eq:Eres_num}
\end{equation}
For the ion-kinetic families ($k_\parallel\simeq k$, $B\simeq 80$~nT), $k\sim 3\times 10^{-4}$--$10^{-2}$~km$^{-1}$ corresponds to $E_{\rm res}$ from a few keV to a few MeV, reproducing the RH and LH peak resonant energies of Section~\ref{sec:generation}. Equation~\eqref{eq:Eres_num} neglects $\omega$ against $\Omega_{ci}$; the neglect holds to $10\%$ at the occurrence peaks and degrades to a factor of two at $k\simeq 10^{-2}$~km$^{-1}$, the upper edge of the band. For the LP-OB family the much longer peak wavelength ($k_\parallel\simeq k\cos 61^\circ\simeq 1.5\times 10^{-5}$~km$^{-1}$ at the peak and the median obliquity) places the dominant cyclotron-resonant energy deep in the relativistic regime ($\gtrsim$100~MeV), well beyond the EPI-Lo range.

\bibliographystyle{aasjournal}
\bibliography{bibTeX_11_11_2022}

@ARTICLE{Dresing25,
       author = {{Dresing}, N. and {Jebaraj}, I.~C. and {Wijsen}, N. and {Palmerio}, E. and {Rodr{\'\i}guez-Garc{\'\i}a}, L. and {Palmroos}, C. and {Gieseler}, J. and {Jarry}, M. and {Asvestari}, E. and {Mitchell}, J.~G. and {Cohen}, C.~M.~S. and {Lee}, C.~O. and {Wei}, W. and {Ramstad}, R. and {Riihonen}, E. and {Oleynik}, P. and {Kouloumvakos}, A. and {Warmuth}, A. and {S{\'a}nchez-Cano}, B. and {Ehresmann}, B. and {Dunn}, P. and {Dudnik}, O. and {Mac Cormack}, C.},
        title = "{On the reason for the widespread energetic storm particle event of 13 March 2023}",
      journal = {\aap},
       volume = {695},
        pages = {A127},
         year = 2025,
        month = feb,
          doi = {10.1051/0004-6361/202453596},
archivePrefix = {arXiv},
       eprint = {2502.06332},
 primaryClass = {astro-ph.SR},
       adsurl = {https://ui.adsabs.harvard.edu/abs/2025arXiv250206332D}
}

@ARTICLE{Wijsen25,
        title = {Freely propagating flanks of wide coronal-mass-ejection-driven shocks: Modelling and observational insights},
        author = {{Wijsen}, Nicolas and {Jebaraj}, Immanuel and {Dresing}, Nina and {Kouloumvakos}, Athanasios and {Palmerio}, Erika and {Rodr\'iguez-Garc\'ia}, Laura and {Lario}, David},
        volume={699},
        pages={A51},
        year={2025},
        journal = {\aap},
        doi = {10.1051/0004-6361/202453598},
        eprint = {2505.02794},
        archiveprefix = {arXiv}
}

@ARTICLE{Jebaraj24b,
       author = {{Jebaraj}, I.~C. and {Agapitov}, O.~V. and {Gedalin}, M. and {Vuorinen}, L. and {Miceli}, M. and {Cohen}, C.~M.~S. and {Voshchepynets}, A. and {Kouloumvakos}, A. and {Dresing}, N. and {Marmyleva}, A. and {Krasnoselskikh}, V. and {Balikhin}, M. and {Mitchell}, J.~G. and {Labrador}, A.~W. and {Wijsen}, N. and {Palmerio}, E. and {Colomban}, L. and {Pomoell}, J. and {Kilpua}, E.~K.~J. and {Pulupa}, M. and {Mozer}, F.~S. and {Raouafi}, N.~E. and {McComas}, D.~J. and {Bale}, S.~D. and {Vainio}, R.},
        title = "{Direct Measurements of Synchrotron-emitting Electrons at Near-Sun Shocks}",
      journal = {\apjl},
         year = 2024,
        month = nov,
       volume = {976},
       number = {1},
          eid = {L7},
        pages = {L7},
          doi = {10.3847/2041-8213/ad8eb8},
archivePrefix = {arXiv},
       eprint = {2410.15933},
 primaryClass = {physics.space-ph},
       adsurl = {https://ui.adsabs.harvard.edu/abs/2024ApJ...976L...7J}
}

@article{Jebaraj24,
	adsurl = {https://ui.adsabs.harvard.edu/abs/2024ApJ...968L...8J},
	author = {{Jebaraj}, Immanuel Christopher and {Agapitov}, Oleksiy and {Krasnoselskikh}, Vladimir and {Vuorinen}, Laura and {Gedalin}, Michael and {Choi}, Kyung-Eun and {Palmerio}, Erika and {Wijsen}, Nicolas and {Dresing}, Nina and {Cohen}, Christina and {Kouloumvakos}, Athanasios and {Balikhin}, Michael and {Vainio}, Rami and {Kilpua}, Emilia and {Afanasiev}, Alexandr and {Verniero}, Jaye and {Mitchell}, John Grant and {Trotta}, Domenico and {Hill}, Matthew and {Raouafi}, Nour and {Bale}, Stuart D.},
	doi = {10.3847/2041-8213/ad4daa},
	eid = {L8},
	journal = {\apjl},
	month = jun,
	number = {1},
	pages = {L8},
	title = {{Acceleration of Electrons and Ions by an ``Almost'' Astrophysical Shock in the Heliosphere}},
	volume = {968},
	year = 2024}

@ARTICLE{Kouloumvakos26,
	author = {{Kouloumvakos}, A. and {Jebaraj}, I.~C. and {Malkov}, M.~A. and {Gedalin}, M. and {Colomban}, L. and {Agapitov}, O.~V.},
	title = {{Wave-Growth-Limited Acceleration in a Near-Parallel Interplanetary Shock Observed by Parker Solar Probe}},
	year = {2026},
	journal = {ApJ Letters, submitted}}

@article{Giacalone26,
	author = {{Giacalone}, J. and {Trotta}, D. and {Mitchell}, D.~G. and {Cohen}, C.~M.~S. and {Fraschetti}, F. and {Bale}, S.~D.},
	title = {{The Source of Energetic Storm Particles for the 2023 March 13 Event Observed by Parker Solar Probe}},
	journal = {\apj},
	year = {2026},
	volume = {996},
	number = {1},
	eid = {87},
	pages = {87},
	doi = {10.3847/1538-4357/ae1ef5},
	adsurl = {https://ui.adsabs.harvard.edu/abs/2026ApJ...996...87G}}

@article{Fox2016,
	adsurl = {https://ui.adsabs.harvard.edu/abs/2016SSRv..204....7F},
	author = {{Fox}, N.~J. and {Velli}, M.~C. and {Bale}, S.~D. and {Decker}, R. and {Driesman}, A. and {Howard}, R.~A. and {Kasper}, J.~C. and {Kinnison}, J. and {Kusterer}, M. and {Lario}, D. and {Lockwood}, M.~K. and {McComas}, D.~J. and {Raouafi}, N.~E. and {Szabo}, A.},
	doi = {10.1007/s11214-015-0211-6},
	journal = {Space Science Reviews},
	month = dec,
	number = {1-4},
	pages = {7-48},
	title = {{The Solar Probe Plus Mission: Humanity's First Visit to Our Star}},
	volume = {204},
	year = 2016}

@article{Pulupa2017,
	adsurl = {https://ui.adsabs.harvard.edu/abs/2017JGRA..122.2836P},
	author = {{Pulupa}, M. and {Bale}, S.~D. and {Bonnell}, J.~W. and {Bowen}, T.~A. and {Carruth}, N. and {Goetz}, K. and {Gordon}, D. and {Harvey}, P.~R. and {Maksimovic}, M. and {Mart{\'\i}nez-Oliveros}, J.~C. and {Moncuquet}, M. and {Saint-Hilaire}, P. and {Seitz}, D. and {Sundkvist}, D.},
	doi = {10.1002/2016JA023345},
	journal = {Journal of Geophysical Research (Space Physics)},
	month = mar,
	number = {3},
	pages = {2836-2854},
	title = {{The Solar Probe Plus Radio Frequency Spectrometer: Measurement requirements, analog design, and digital signal processing}},
	volume = {122},
	year = 2017}

@article{Sonnerup98,
	adsurl = {https://ui.adsabs.harvard.edu/abs/1998ISSIR...1..185S},
	author = {{Sonnerup}, Bengt U. {\"O}. and {Scheible}, Maureen},
	journal = {ISSI Scientific Reports Series},
	month = jan,
	pages = {185-220},
	title = {{Minimum and Maximum Variance Analysis}},
	volume = {1},
	year = 1998}

@article{Abraham72,
	adsurl = {https://ui.adsabs.harvard.edu/abs/1972JGR....77..736A},
	author = {{Abraham-Shrauner}, Barbara},
	doi = {10.1029/JA077i004p00736},
	journal = {Journal of Geophysical Research},
	month = jan,
	number = {4},
	pages = {736},
	title = {{Determination of magnetohydrodynamic shock normals}},
	volume = {77},
	year = 1972}

@article{Caprioli14,
	adsurl = {https://ui.adsabs.harvard.edu/abs/2014ApJ...783...91C},
	archiveprefix = {arXiv},
	author = {{Caprioli}, D. and {Spitkovsky}, A.},
	doi = {10.1088/0004-637X/783/2/91},
	eid = {91},
	eprint = {1310.2943},
	journal = {The Astrophysical Journal},
	month = mar,
	number = {2},
	pages = {91},
	primaryclass = {astro-ph.HE},
	title = {{Simulations of Ion Acceleration at Non-relativistic Shocks. I. Acceleration Efficiency}},
	volume = {783},
	year = 2014}

@article{Vainio03,
	adsurl = {https://ui.adsabs.harvard.edu/abs/2003A&A...406..735V},
	author = {{Vainio}, R.},
	doi = {10.1051/0004-6361:20030822},
	journal = {Astronomy and Astrophysics},
	month = aug,
	pages = {735-740},
	title = {{On the generation of Alfv{\'e}n waves by solar energetic particles}},
	volume = {406},
	year = 2003}

@article{Kennel85,
	adsurl = {https://ui.adsabs.harvard.edu/abs/1985GMS....34....1K},
	author = {{Kennel}, C.~F. and {Edmiston}, J.~P. and {Hada}, T.},
	doi = {10.1029/GM034p0001},
	journal = {Geophysical Monograph Series},
	month = jan,
	pages = {1-36},
	title = {{A quarter century of collisionless shock research}},
	volume = {34},
	year = 1985}

@article{Zhao21,
	adsurl = {https://ui.adsabs.harvard.edu/abs/2021A&A...656A...3Z},
	archiveprefix = {arXiv},
	author = {{Zhao}, L. -L. and {Zank}, G.~P. and {He}, J.~S. and {Telloni}, D. and {Hu}, Q. and {Li}, G. and {Nakanotani}, M. and {Adhikari}, L. and {Kilpua}, E.~K.~J. and {Horbury}, T.~S. and {O'Brien}, H. and {Evans}, V. and {Angelini}, V.},
	doi = {10.1051/0004-6361/202140450},
	eid = {A3},
	eprint = {2102.03301},
	journal = {Astronomy and Astrophysics},
	month = dec,
	pages = {A3},
	primaryclass = {physics.space-ph},
	title = {{Turbulence and wave transmission at an ICME-driven shock observed by the Solar Orbiter and Wind}},
	volume = {656},
	year = 2021}

@article{Lembege04,
	adsurl = {https://ui.adsabs.harvard.edu/abs/2004SSRv..110..161L},
	author = {{Lembege}, B. and {Giacalone}, J. and {Scholer}, M. and {Hada}, T. and {Hoshino}, M. and {Krasnoselskikh}, V. and {Kucharek}, H. and {Savoini}, P. and {Terasawa}, T.},
	doi = {10.1023/B:SPAC.0000023372.12232.b7},
	journal = {Space Science Reviews},
	month = jan,
	number = {3},
	pages = {161-226},
	title = {{Selected Problems in Collisionless-Shock Physics}},
	volume = {110},
	year = 2004}

@ARTICLE{Drury83,
       author = {{Drury}, L. Oc.},
        title = "{REVIEW ARTICLE: An introduction to the theory of diffusive shock acceleration of energetic particles in tenuous plasmas}",
      journal = {Reports on Progress in Physics},
         year = 1983,
        month = aug,
       volume = {46},
       number = {8},
        pages = {973-1027},
          doi = {10.1088/0034-4885/46/8/002},
       adsurl = {https://ui.adsabs.harvard.edu/abs/1983RPPh...46..973D}
}

@article{Krasnoselskikh13,
	adsurl = {https://ui.adsabs.harvard.edu/abs/2013SSRv..178..535K},
	archiveprefix = {arXiv},
	author = {{Krasnoselskikh}, V. and {Balikhin}, M. and {Walker}, S.~N. and {Schwartz}, S. and {Sundkvist}, D. and {Lobzin}, V. and {Gedalin}, M. and {Bale}, S.~D. and {Mozer}, F. and {Soucek}, J. and {Hobara}, Y. and {Comisel}, H.},
	doi = {10.1007/s11214-013-9972-y},
	eprint = {1303.0190},
	journal = {Space Science Reviews},
	month = oct,
	number = {2-4},
	pages = {535-598},
	primaryclass = {physics.space-ph},
	title = {{The Dynamic Quasiperpendicular Shock: Cluster Discoveries}},
	volume = {178},
	year = 2013}

@article{Vainio00,
	adsurl = {https://ui.adsabs.harvard.edu/abs/2000ApJS..131..519V},
	author = {{Vainio}, R.},
	doi = {10.1086/317372},
	journal = {The Astrophysical Journal Supplement Series},
	month = dec,
	number = {2},
	pages = {519-529},
	title = {{Charged-Particle Resonance Conditions and Transport Coefficients in Slab-Mode Waves}},
	volume = {131},
	year = 2000}

@inproceedings{Axford77,
	adsurl = {https://ui.adsabs.harvard.edu/abs/1977ICRC...11..132A},
	author = {{Axford}, W.~I. and {Leer}, E. and {Skadron}, G.},
	booktitle = {International Cosmic Ray Conference},
	month = jan,
	pages = {132},
	series = {International Cosmic Ray Conference},
	title = {{The Acceleration of Cosmic Rays by Shock Waves}},
	volume = {11},
	year = 1977}

@article{Bell78,
	adsurl = {https://ui.adsabs.harvard.edu/abs/1978MNRAS.182..147B},
	author = {{Bell}, A.~R.},
	doi = {10.1093/mnras/182.2.147},
	journal = {Monthly Notices of the Royal Astronomical Society},
	month = jan,
	pages = {147-156},
	title = {{The acceleration of cosmic rays in shock fronts - I.}},
	volume = {182},
	year = 1978}

@article{Krymskii77,
	adsurl = {https://ui.adsabs.harvard.edu/abs/1977DoSSR.234.1306K},
	author = {{Krymskii}, G.~F.},
	journal = {Akademiia Nauk SSSR Doklady},
	month = jun,
	pages = {1306-1308},
	title = {{A regular mechanism for the acceleration of charged particles on the front of a shock wave}},
	volume = {234},
	year = 1977}

@article{Blandford78,
	adsurl = {https://ui.adsabs.harvard.edu/abs/1978ApJ...221L..29B},
	author = {{Blandford}, R.~D. and {Ostriker}, J.~P.},
	doi = {10.1086/182658},
	journal = {The Astrophysical Journal Letters},
	month = apr,
	pages = {L29-L32},
	title = {{Particle acceleration by astrophysical shocks.}},
	volume = {221},
	year = 1978}

@article{Lee83,
	adsurl = {https://ui.adsabs.harvard.edu/abs/1983JGR....88.6109L},
	author = {{Lee}, M.~A.},
	doi = {10.1029/JA088iA08p06109},
	journal = {Journal of Geophysical Research},
	month = aug,
	number = {A8},
	pages = {6109-6120},
	title = {{Coupled hydromagnetic wave excitation and ion acceleration at interplanetary traveling shocks}},
	volume = {88},
	year = 1983}

@article{Malkov01,
	adsurl = {https://ui.adsabs.harvard.edu/abs/2001RPPh...64..429M},
	author = {{Malkov}, M.~A. and {Drury}, L. O'C.},
	doi = {10.1088/0034-4885/64/4/201},
	journal = {Reports on Progress in Physics},
	month = apr,
	number = {4},
	pages = {429-481},
	title = {{Nonlinear theory of diffusive acceleration of particles by shock waves}},
	volume = {64},
	year = 2001}

@article{Vainio14,
	author = {Vainio, Rami and P{\"o}nni, Arttu and Battarbee, Markus and Koskinen, Hannu EJ and Afanasiev, Alexandr and Laitinen, Timo Lauri Mikael},
	journal = {Journal of Space Weather and Space Climate},
	pages = {A08},
	publisher = {EDP sciences},
	title = {A semi-analytical foreshock model for energetic storm particle events inside 1 AU},
	volume = {4},
	year = {2014}}

@article{Lagage83,
	adsurl = {https://ui.adsabs.harvard.edu/abs/1983A&A...118..223L},
	author = {{Lagage}, P.~O. and {Cesarsky}, C.~J.},
	journal = {Astronomy and Astrophysics},
	month = feb,
	number = {2},
	pages = {223-228},
	title = {{Cosmic-ray shock acceleration in the presence of self-excited waves}},
	volume = {118},
	year = 1983}

@article{Kennel86,
	adsurl = {https://ui.adsabs.harvard.edu/abs/1986JGR....9111917K},
	author = {{Kennel}, C.~F. and {Coroniti}, F.~V. and {Scarf}, F.~L. and {Livesey}, W.~A. and {Russell}, C.~T. and {Smith}, E.~J. and {Wenzel}, K.~P. and {Scholer}, M.},
	doi = {10.1029/JA091iA11p11917},
	journal = {Journal of Geophysical Research},
	month = nov,
	number = {A11},
	pages = {11917-11928},
	title = {{A test of Lee's quasi-linear theory of ion acceleration by interplanetary traveling shocks}},
	volume = {91},
	year = 1986}

@article{Sundkvist12,
	adsurl = {https://ui.adsabs.harvard.edu/abs/2012PhRvL.108b5002S},
	author = {{Sundkvist}, David and {Krasnoselskikh}, V. and {Bale}, S.~D. and {Schwartz}, S.~J. and {Soucek}, J. and {Mozer}, F.},
	doi = {10.1103/PhysRevLett.108.025002},
	eid = {025002},
	journal = {Physical Review Letters},
	month = jan,
	number = {2},
	pages = {025002},
	title = {{Dispersive Nature of High Mach Number Collisionless Plasma Shocks: Poynting Flux of Oblique Whistler Waves}},
	volume = {108},
	year = 2012}

@article{Burgess05,
	adsurl = {https://ui.adsabs.harvard.edu/abs/2005SSRv..118..205B},
	author = {{Burgess}, D. and {Lucek}, E.~A. and {Scholer}, M. and {Bale}, S.~D. and {Balikhin}, M.~A. and {Balogh}, A. and {Horbury}, T.~S. and {Krasnoselskikh}, V.~V. and {Kucharek}, H. and {Lemb{\`e}ge}, B. and {M{\"o}bius}, E. and {Schwartz}, S.~J. and {Thomsen}, M.~F. and {Walker}, S.~N.},
	doi = {10.1007/s11214-005-3832-3},
	journal = {\ssr},
	month = jun,
	number = {1-4},
	pages = {205-222},
	title = {{Quasi-parallel Shock Structure and Processes}},
	volume = {118},
	year = 2005}

@article{Malkov98b,
	adsurl = {https://ui.adsabs.harvard.edu/abs/1998PhRvE..58.4911M},
	archiveprefix = {arXiv},
	author = {{Malkov}, M.~A.},
	doi = {10.1103/PhysRevE.58.4911},
	eprint = {astro-ph/9806340},
	journal = {Physical Review E},
	month = oct,
	number = {4},
	pages = {4911-4928},
	primaryclass = {astro-ph},
	title = {{Ion leakage from quasiparallel collisionless shocks: Implications for injection and shock dissipation}},
	volume = {58},
	year = 1998}

@article{Livi22,
	adsurl = {https://ui.adsabs.harvard.edu/abs/2022ApJ...938..138L},
	author = {{Livi}, Roberto and {Larson}, Davin E. and {Kasper}, Justin C. and {Abiad}, Robert and {Case}, A.~W. and {Klein}, Kristopher G. and {Curtis}, David W. and {Dalton}, Gregory and {Stevens}, Michael and {Korreck}, Kelly E. and {Ho}, George and {Robinson}, Miles and {Tiu}, Chris and {Whittlesey}, Phyllis L. and {Verniero}, Jaye L. and {Halekas}, Jasper and {McFadden}, James and {Marckwordt}, Mario and {Slagle}, Amanda and {Abatcha}, Mamuda and {Rahmati}, Ali and {McManus}, Michael D.},
	doi = {10.3847/1538-4357/ac93f5},
	eid = {138},
	journal = {The Astrophysical Journal},
	month = oct,
	number = {2},
	pages = {138},
	title = {{The Solar Probe ANalyzer-Ions on the Parker Solar Probe}},
	volume = {938},
	year = 2022}

@ARTICLE{Raptis26,
       author = {{Raptis}, Savvas and {Trotta}, Domenico and {Turner}, Drew L. and {Blanco-Cano}, X{\'o}chitl and {Hietala}, Heli and {Karlsson}, Tomas and {Jebaraj}, Immanuel Christopher and {Vasko}, Ivan Y. and {Osmane}, Adnane and {Takahashi}, Kazue and {Lario}, David and {Wilson}, III, Lynn B. and {Howes}, Gregory G. and {Wimmer-Schweingruber}, Robert F.},
        title = "{Compressive Structures in the Foreshock of Collisionless Shocks}",
      journal = {\apjl},
         year = 2026,
        month = apr,
       volume = {1000},
       number = {2},
          eid = {L55},
        pages = {L55},
          doi = {10.3847/2041-8213/ae53e5},
archivePrefix = {arXiv},
       eprint = {2603.17882},
 primaryClass = {physics.space-ph},
       adsurl = {https://ui.adsabs.harvard.edu/abs/2026ApJ..1000L..55R}
}

@ARTICLE{chust2021,
       author = {{Chust}, T. and {Kretzschmar}, M. and {Graham}, D.~B. and {Le Contel}, O. and {Retin{\`o}}, A. and {Alexandrova}, A. and {Berthomier}, M. and {Hadid}, L.~Z. and {Sahraoui}, F. and {Jeandet}, A. and {Leroy}, P. and {Pellion}, J.-C. and {Bouzid}, V. and {Katra}, B. and {Piberne}, R. and {Khotyaintsev}, Yu. V. and {Vaivads}, A. and {Krasnoselskikh}, V. and {Sou{\v{c}}ek}, J. and {Santol{\'\i}k}, O. and {Lorf{\`e}vre}, E. and {Plettemeier}, D. and {Steller}, M. and {{\v{S}}tver{\'a}k}, {\v{S}}. and {Tr{\'a}vn{\'\i}{\v{c}}ek}, P. and {Vecchio}, A. and {Maksimovic}, M. and {Bale}, S.~D. and {Horbury}, T.~S. and {O'Brien}, H. and {Evans}, V. and {Angelini}, V.},
        title = "{Observations of whistler mode waves by Solar Orbiter's RPW Low Frequency Receiver (LFR): In-flight performance and first results}",
      journal = {\aap},
         year = 2021,
        month = dec,
       volume = {656},
          eid = {A17},
        pages = {A17},
          doi = {10.1051/0004-6361/202140932},
       adsurl = {https://ui.adsabs.harvard.edu/abs/2021A&A...656A..17C}
}

@ARTICLE{colomban2025,
       author = {{Colomban}, L. and {Agapitov}, O.~V. and {Krasnoselskikh}, V. and {Choi}, K.-E. and {Kretzschmar}, M. and {Dudok de Wit}, T. and {Mozer}, F.~S. and {Bonnell}, J.~W. and {Bale}, S. and {Malaspina}, D. and {Raouafi}, N.~E. and {Pulupa}, M.},
        title = "{Polarization Properties of Whistler Waves From the First 17 Parker Solar Probe Encounters}",
      journal = {\grl},
         year = 2025,
        month = may,
       volume = {52},
       number = {10},
          eid = {e2025GL114622},
        pages = {e2025GL114622},
          doi = {10.1029/2025GL114622},
       adsurl = {https://ui.adsabs.harvard.edu/abs/2025GeoRL..5214622C}
}

@ARTICLE{Mozer2020,
       author = {{Mozer}, F.~S. and {Agapitov}, O.~V. and {Bale}, S.~D. and {Bonnell}, J.~W. and {Bowen}, T.~A. and {Vasko}, I.},
        title = "{DC and Low-Frequency Electric Field Measurements on the Parker Solar Probe}",
      journal = {Journal of Geophysical Research (Space Physics)},
         year = 2020,
        month = sep,
       volume = {125},
       number = {9},
          eid = {e27980},
        pages = {e27980},
          doi = {10.1029/2020JA027980},
       adsurl = {https://ui.adsabs.harvard.edu/abs/2020JGRA..12527980M}
}

@ARTICLE{colomban2024,
       author = {{Colomban}, L. and {Kretzschmar}, M. and {Krasnoselkikh}, V. and {Agapitov}, O.~V. and {Froment}, C. and {Maksimovic}, M. and {Berthomier}, M. and {Khotyaintsev}, Yu. V. and {Graham}, D.~B. and {Bale}, S.},
        title = "{Quantifying the diffusion of suprathermal electrons by whistler waves between 0.2 and 1 AU with Solar Orbiter and Parker Solar Probe}",
      journal = {\aap},
         year = 2024,
        month = apr,
       volume = {684},
          eid = {A143},
        pages = {A143},
          doi = {10.1051/0004-6361/202347489},
archivePrefix = {arXiv},
       eprint = {2402.06016},
 primaryClass = {astro-ph.SR},
       adsurl = {https://ui.adsabs.harvard.edu/abs/2024A&A...684A.143C}
}

@article{Bale16,
	adsurl = {https://ui.adsabs.harvard.edu/abs/2016SSRv..204...49B},
	author = {{Bale}, S.~D. and {Goetz}, K. and {Harvey}, P.~R. and {Turin}, P. and {Bonnell}, J.~W. and {Dudok de Wit}, T. and {Ergun}, R.~E. and {MacDowall}, R.~J. and {Pulupa}, M. and {Andre}, M. and {Bolton}, M. and {Bougeret}, J. -L. and {Bowen}, T.~A. and {Burgess}, D. and {Cattell}, C.~A. and {Chandran}, B.~D.~G. and {Chaston}, C.~C. and {Chen}, C.~H.~K. and {Choi}, M.~K. and {Connerney}, J.~E. and {Cranmer}, S. and {Diaz-Aguado}, M. and {Donakowski}, W. and {Drake}, J.~F. and {Farrell}, W.~M. and {Fergeau}, P. and {Fermin}, J. and {Fischer}, J. and {Fox}, N. and {Glaser}, D. and {Goldstein}, M. and {Gordon}, D. and {Hanson}, E. and {Harris}, S.~E. and {Hayes}, L.~M. and {Hinze}, J.~J. and {Hollweg}, J.~V. and {Horbury}, T.~S. and {Howard}, R.~A. and {Hoxie}, V. and {Jannet}, G. and {Karlsson}, M. and {Kasper}, J.~C. and {Kellogg}, P.~J. and {Kien}, M. and {Klimchuk}, J.~A. and {Krasnoselskikh}, V.~V. and {Krucker}, S. and {Lynch}, J.~J. and {Maksimovic}, M. and {Malaspina}, D.~M. and {Marker}, S. and {Martin}, P. and {Martinez-Oliveros}, J. and {McCauley}, J. and {McComas}, D.~J. and {McDonald}, T. and {Meyer-Vernet}, N. and {Moncuquet}, M. and {Monson}, S.~J. and {Mozer}, F.~S. and {Murphy}, S.~D. and {Odom}, J. and {Oliverson}, R. and {Olson}, J. and {Parker}, E.~N. and {Pankow}, D. and {Phan}, T. and {Quataert}, E. and {Quinn}, T. and {Ruplin}, S.~W. and {Salem}, C. and {Seitz}, D. and {Sheppard}, D.~A. and {Siy}, A. and {Stevens}, K. and {Summers}, D. and {Szabo}, A. and {Timofeeva}, M. and {Vaivads}, A. and {Velli}, M. and {Yehle}, A. and {Werthimer}, D. and {Wygant}, J.~R.},
	doi = {10.1007/s11214-016-0244-5},
	journal = {Space Science Reviews},
	month = dec,
	number = {1-4},
	pages = {49-82},
	title = {{The FIELDS Instrument Suite for Solar Probe Plus. Measuring the Coronal Plasma and Magnetic Field, Plasma Waves and Turbulence, and Radio Signatures of Solar Transients}},
	volume = {204},
	year = 2016}

@ARTICLE{Malkov06,
       author = {{Malkov}, M.~A. and {Diamond}, P.~H.},
        title = "{Nonlinear Shock Acceleration beyond the Bohm Limit}",
      journal = {\apj},
         year = 2006,
        month = may,
       volume = {642},
       number = {1},
        pages = {244-259},
          doi = {10.1086/500445},
archivePrefix = {arXiv},
       eprint = {astro-ph/0509235},
 primaryClass = {astro-ph},
       adsurl = {https://ui.adsabs.harvard.edu/abs/2006ApJ...642..244M}
}

@article{Kasper16,
	adsurl = {https://ui.adsabs.harvard.edu/abs/2016SSRv..204..131K},
	author = {{Kasper}, Justin C. and {Abiad}, Robert and {Austin}, Gerry and {Balat-Pichelin}, Marianne and {Bale}, Stuart D. and {Belcher}, John W. and {Berg}, Peter and {Bergner}, Henry and {Berthomier}, Matthieu and {Bookbinder}, Jay and {Brodu}, Etienne and {Caldwell}, David and {Case}, Anthony W. and {Chandran}, Benjamin D.~G. and {Cheimets}, Peter and {Cirtain}, Jonathan W. and {Cranmer}, Steven R. and {Curtis}, David W. and {Daigneau}, Peter and {Dalton}, Greg and {Dasgupta}, Brahmananda and {DeTomaso}, David and {Diaz-Aguado}, Millan and {Djordjevic}, Blagoje and {Donaskowski}, Bill and {Effinger}, Michael and {Florinski}, Vladimir and {Fox}, Nichola and {Freeman}, Mark and {Gallagher}, Dennis and {Gary}, S. Peter and {Gauron}, Tom and {Gates}, Richard and {Goldstein}, Melvin and {Golub}, Leon and {Gordon}, Dorothy A. and {Gurnee}, Reid and {Guth}, Giora and {Halekas}, Jasper and {Hatch}, Ken and {Heerikuisen}, Jacob and {Ho}, George and {Hu}, Qiang and {Johnson}, Greg and {Jordan}, Steven P. and {Korreck}, Kelly E. and {Larson}, Davin and {Lazarus}, Alan J. and {Li}, Gang and {Livi}, Roberto and {Ludlam}, Michael and {Maksimovic}, Milan and {McFadden}, James P. and {Marchant}, William and {Maruca}, Bennet A. and {McComas}, David J. and {Messina}, Luciana and {Mercer}, Tony and {Park}, Sang and {Peddie}, Andrew M. and {Pogorelov}, Nikolai and {Reinhart}, Matthew J. and {Richardson}, John D. and {Robinson}, Miles and {Rosen}, Irene and {Skoug}, Ruth M. and {Slagle}, Amanda and {Steinberg}, John T. and {Stevens}, Michael L. and {Szabo}, Adam and {Taylor}, Ellen R. and {Tiu}, Chris and {Turin}, Paul and {Velli}, Marco and {Webb}, Gary and {Whittlesey}, Phyllis and {Wright}, Ken and {Wu}, S.~T. and {Zank}, Gary},
	doi = {10.1007/s11214-015-0206-3},
	journal = {Space Science Reviews},
	month = dec,
	number = {1-4},
	pages = {131-186},
	title = {{Solar Wind Electrons Alphas and Protons (SWEAP) Investigation: Design of the Solar Wind and Coronal Plasma Instrument Suite for Solar Probe Plus}},
	volume = {204},
	year = 2016}

@article{McComas16,
	adsurl = {https://ui.adsabs.harvard.edu/abs/2016SSRv..204..187M},
	author = {{McComas}, D.~J. and {Alexander}, N. and {Angold}, N. and {Bale}, S. and {Beebe}, C. and {Birdwell}, B. and {Boyle}, M. and {Burgum}, J.~M. and {Burnham}, J.~A. and {Christian}, E.~R. and {Cook}, W.~R. and {Cooper}, S.~A. and {Cummings}, A.~C. and {Davis}, A.~J. and {Desai}, M.~I. and {Dickinson}, J. and {Dirks}, G. and {Do}, D.~H. and {Fox}, N. and {Giacalone}, J. and {Gold}, R.~E. and {Gurnee}, R.~S. and {Hayes}, J.~R. and {Hill}, M.~E. and {Kasper}, J.~C. and {Kecman}, B. and {Klemic}, J. and {Krimigis}, S.~M. and {Labrador}, A.~W. and {Layman}, R.~S. and {Leske}, R.~A. and {Livi}, S. and {Matthaeus}, W.~H. and {McNutt}, R.~L. and {Mewaldt}, R.~A. and {Mitchell}, D.~G. and {Nelson}, K.~S. and {Parker}, C. and {Rankin}, J.~S. and {Roelof}, E.~C. and {Schwadron}, N.~A. and {Seifert}, H. and {Shuman}, S. and {Stokes}, M.~R. and {Stone}, E.~C. and {Vandegriff}, J.~D. and {Velli}, M. and {von Rosenvinge}, T.~T. and {Weidner}, S.~E. and {Wiedenbeck}, M.~E. and {Wilson}, P.},
	doi = {10.1007/s11214-014-0059-1},
	journal = {Space Science Reviews},
	month = dec,
	number = {1-4},
	pages = {187-256},
	title = {{Integrated Science Investigation of the Sun (ISIS): Design of the Energetic Particle Investigation}},
	volume = {204},
	year = 2016}

@article{Hill17,
	adsurl = {https://ui.adsabs.harvard.edu/abs/2017JGRA..122.1513H},
	author = {{Hill}, M.~E. and {Mitchell}, D.~G. and {Andrews}, G.~B. and {Cooper}, S.~A. and {Gurnee}, R.~S. and {Hayes}, J.~R. and {Layman}, R.~S. and {McNutt}, R.~L. and {Nelson}, K.~S. and {Parker}, C.~W. and {Schlemm}, C.~E. and {Stokes}, M.~R. and {Begley}, S.~M. and {Boyle}, M.~P. and {Burgum}, J.~M. and {Do}, D.~H. and {Dupont}, A.~R. and {Gold}, R.~E. and {Haggerty}, D.~K. and {Hoffer}, E.~M. and {Hutcheson}, J.~C. and {Jaskulek}, S.~E. and {Krimigis}, S.~M. and {Liang}, S.~X. and {London}, S.~M. and {Noble}, M.~W. and {Roelof}, E.~C. and {Seifert}, H. and {Strohbehn}, K. and {Vandegriff}, J.~D. and {Westlake}, J.~H.},
	doi = {10.1002/2016JA022614},
	journal = {Journal of Geophysical Research (Space Physics)},
	month = feb,
	number = {2},
	pages = {1513-1530},
	title = {{The Mushroom: A half-sky energetic ion and electron detector}},
	volume = {122},
	year = 2017}

@article{Kennel81,
	adsurl = {https://ui.adsabs.harvard.edu/abs/1981JGR....86.4325K},
	author = {{Kennel}, C.~F.},
	doi = {10.1029/JA086iA06p04325},
	journal = {Journal of Geophysical Research},
	month = jun,
	number = {A6},
	pages = {4325-4330},
	title = {{Collisionless shocks and upstream waves and particles: Introductory remarks}},
	volume = {86},
	year = 1981}

@article{Lario15,
	adsurl = {https://ui.adsabs.harvard.edu/abs/2015ApJ...813...85L},
	archiveprefix = {arXiv},
	author = {{Lario}, D. and {Decker}, R.~B. and {Roelof}, E.~C. and {Vi{\~n}as}, A. -F.},
	doi = {10.1088/0004-637X/813/2/85},
	eid = {85},
	eprint = {1509.04368},
	journal = {The Astrophysical Journal},
	month = nov,
	number = {2},
	pages = {85},
	primaryclass = {physics.space-ph},
	title = {{Energetic Particle Pressure at Interplanetary Shocks: STEREO-A Observations}},
	volume = {813},
	year = 2015}

@article{Drury81,
	adsurl = {https://ui.adsabs.harvard.edu/abs/1981ApJ...248..344D},
	author = {{Drury}, L. O'C. and {Voelk}, H.~J.},
	doi = {10.1086/159159},
	journal = {The Astrophysical Journal},
	month = aug,
	pages = {344-351},
	title = {{Hydromagnetic shock structure in the presence of cosmic rays}},
	volume = {248},
	year = 1981}

@article{Karbashewski23,
	adsurl = {https://ui.adsabs.harvard.edu/abs/2023ApJ...947...73K},
	archiveprefix = {arXiv},
	author = {{Karbashewski}, S. and {Agapitov}, O.~V. and {Kim}, H.~Y. and {Mozer}, F.~S. and {Bonnell}, J.~W. and {Froment}, C. and {de Wit}, T. Dudok and {Bale}, Stuart D. and {Malaspina}, D. and {Raouafi}, N.~E.},
	doi = {10.3847/1538-4357/acc527},
	eid = {73},
	eprint = {2304.01185},
	journal = {The Astrophysical Journal},
	month = apr,
	number = {2},
	pages = {73},
	primaryclass = {physics.space-ph},
	title = {{Whistler Wave Observations by Parker Solar Probe During Encounter 1: Counter-propagating Whistlers Collocated with Magnetic Field Inhomogeneities and their Application to Electric Field Measurement Calibration}},
	volume = {947},
	year = 2023}

@article{Colburn66,
	adsurl = {https://ui.adsabs.harvard.edu/abs/1966SSRv....5..439C},
	author = {{Colburn}, D.~S. and {Sonett}, C.~P.},
	doi = {10.1007/BF00240575},
	journal = {Space Science Reviews},
	month = jun,
	number = {4},
	pages = {439-506},
	title = {{Discontinuities in the Solar Wind}},
	volume = {5},
	year = 1966}

@article{Gedalin98,
	adsurl = {https://ui.adsabs.harvard.edu/abs/1998PhPl....5..127G},
	author = {{Gedalin}, M.},
	doi = {10.1063/1.872681},
	journal = {Physics of Plasmas},
	month = jan,
	number = {1},
	pages = {127-132},
	title = {{Low-frequency nonlinear stationary waves and fast shocks: Hydrodynamical description}},
	volume = {5},
	year = 1998}

@article{Schwartz91,
	adsurl = {https://ui.adsabs.harvard.edu/abs/1991GeoRL..18..373S},
	author = {{Schwartz}, Steven J. and {Burgess}, David},
	doi = {10.1029/91GL00138},
	journal = {Geophysical Research Letters},
	month = mar,
	number = {3},
	pages = {373-376},
	title = {{Quasi-parallel shocks: A patchwork of three-dimensional structures}},
	volume = {18},
	year = 1991}

@ARTICLE{Haggerty20,
       author = {{Haggerty}, Colby C. and {Caprioli}, Damiano},
        title = "{Kinetic Simulations of Cosmic-Ray-modified Shocks. I. Hydrodynamics}",
      journal = {\apj},
         year = 2020,
        month = dec,
       volume = {905},
       number = {1},
          eid = {1},
        pages = {1},
          doi = {10.3847/1538-4357/abbe06},
archivePrefix = {arXiv},
       eprint = {2008.12308},
 primaryClass = {astro-ph.HE},
       adsurl = {https://ui.adsabs.harvard.edu/abs/2020ApJ...905....1H}
}

@ARTICLE{Caprioli20,
       author = {{Caprioli}, Damiano and {Haggerty}, Colby C. and {Blasi}, Pasquale},
        title = "{Kinetic Simulations of Cosmic-Ray-modified Shocks. II. Particle Spectra}",
      journal = {\apj},
         year = 2020,
        month = dec,
       volume = {905},
       number = {1},
          eid = {2},
        pages = {2},
          doi = {10.3847/1538-4357/abbe05},
archivePrefix = {arXiv},
       eprint = {2009.00007},
 primaryClass = {astro-ph.HE},
       adsurl = {https://ui.adsabs.harvard.edu/abs/2020ApJ...905....2C}
}

@incollection{Bykov19book,
	adsurl = {https://ui.adsabs.harvard.edu/abs/2019supe.book..419B},
	author = {{Bykov}, A.~M. and {Ellison}, D.~C. and {Marcowith}, A. and {Osipov}, S.~M.},
	booktitle = {Supernovae. Series: Space Sciences Series of ISSI},
	doi = {10.1007/978-94-024-1581-0_15},
	editor = {{Bykov}, Andrei and {Roger}, Chevalier and {Raymond}, John and {Thielemann}, Friedrich-Karl and {Falanga}, Maurizio and {von Steiger}, Rudolf},
	pages = {419-452},
	publisher = {Springer},
	title = {{Cosmic Ray Production in Supernovae}},
	volume = {68},
	year = 2019}

@article{Borovsky19,
	author = {Borovsky, Joseph E.},
	doi = {https://doi.org/10.1029/2019JA027518},
	eprint = {https://agupubs.onlinelibrary.wiley.com/doi/pdf/10.1029/2019JA027518},
	journal = {Journal of Geophysical Research (Space Physics)},
	note = {e2019JA027518 2019JA027518},
	number = {6},
	pages = {e2019JA027518},
	title = {A Statistical Analysis of the Fluctuations in the Upstream and Downstream Plasmas of 109 Strong-Compression Interplanetary Shocks at 1AU},
	url = {https://agupubs.onlinelibrary.wiley.com/doi/abs/10.1029/2019JA027518},
	volume = {125},
	year = {2020}}

@article{Schlickeiser98,
  author = {{Schlickeiser}, R. and {Miller}, J.~A.},
  title = {{Quasi-linear Theory of Cosmic Ray Transport and Acceleration: The Role of Oblique Magnetohydrodynamic Waves and Transit-Time Damping}},
  journal = {\apj},
  volume = {492},
  pages = {352},
  year = {1998},
  doi = {10.1086/305023},
  adsurl = {https://ui.adsabs.harvard.edu/abs/1998ApJ...492..352S}
}

@article{Malkov26,
  author = {{Malkov}, M.~A. and {Jebaraj}, I.~C.},
  title = {{Magnetic Pumping: From Plasma Heating to Particle Acceleration}},
  journal = {arXiv e-prints},
  eprint = {2601.09807},
  primaryclass = {astro-ph.HE},
  year = {2026},
  month = jan,
  doi = {10.48550/arXiv.2601.09807}
}

@article{Means72,
  author = {{Means}, J. D.},
  title = {{Use of the three-dimensional covariance matrix in analyzing the polarization properties of plane waves}},
  journal = {Journal of Geophysical Research},
  volume = {77},
  number = {28},
  pages = {5551--5559},
  year = {1972},
  doi = {10.1029/JA077i028p05551}
}

@article{Santolik03,
  author = {{Santol{\'\i}k}, O. and {Parrot}, M. and {Lefeuvre}, F.},
  title = {{Singular value decomposition methods for wave propagation analysis}},
  journal = {Radio Science},
  volume = {38},
  number = {1},
  pages = {1010},
  year = {2003},
  doi = {10.1029/2000RS002523}
}

@article{Taubenschuss19,
  author = {{Taubenschuss}, U. and {Santol{\'\i}k}, O.},
  title = {{Wave Polarization Analyzed by Singular Value Decomposition of the Spectral Matrix in the Presence of Noise}},
  journal = {Surveys in Geophysics},
  volume = {40},
  pages = {39--69},
  year = {2019},
  doi = {10.1007/s10712-018-9496-9}
}

@article{DruryFalle86,
  author = {{Drury}, L. O'C. and {Falle}, S. A. E. G.},
  title = {{On the stability of shocks modified by particle acceleration}},
  journal = {Monthly Notices of the Royal Astronomical Society},
  volume = {223},
  pages = {353--376},
  year = {1986},
  doi = {10.1093/mnras/223.2.353}
}

@article{ZankAxfordMcKenzie90,
  author = {{Zank}, G. P. and {Axford}, W. I. and {McKenzie}, J. F.},
  title = {{Instabilities in energetic-particle-modified shocks}},
  journal = {Astronomy and Astrophysics},
  volume = {233},
  pages = {275--284},
  year = {1990}
}

@article{KulsrudPearce69,
  author = {{Kulsrud}, R. M. and {Pearce}, W. P.},
  title = {{The Effect of Wave-Particle Interactions on the Propagation of Cosmic Rays}},
  journal = {The Astrophysical Journal},
  volume = {156},
  pages = {445--469},
  year = {1969},
  doi = {10.1086/149981}
}

@article{Eastwood05,
  author = {{Eastwood}, J. P. and {Lucek}, E. A. and {Mazelle}, C. and {Meziane}, K. and {Narita}, Y. and {Pickett}, J. and {Treumann}, R. A.},
  title = {{The Foreshock}},
  journal = {Space Science Reviews},
  volume = {118},
  pages = {41--94},
  year = {2005},
  doi = {10.1007/s11214-005-3824-3}
}

@article{HoppeRussell83,
  author = {{Hoppe}, M. M. and {Russell}, C. T.},
  title = {{Plasma rest frame frequencies and polarizations of the low-frequency upstream waves: ISEE 1 and 2 observations}},
  journal = {Journal of Geophysical Research},
  volume = {88},
  pages = {2021--2027},
  year = {1983},
  doi = {10.1029/JA088iA03p02021}
}

@article{HadaKennelTerasawa87,
  author = {{Hada}, T. and {Kennel}, C. F. and {Terasawa}, T.},
  title = {{Excitation of compressional waves and the formation of shocklets in the Earth's foreshock}},
  journal = {Journal of Geophysical Research},
  volume = {92},
  pages = {4423--4435},
  year = {1987},
  doi = {10.1029/JA092iA05p04423}
}

@article{Gary85,
  author = {{Gary}, S. P.},
  title = {{Electromagnetic ion beam instabilities: Hot beams at interplanetary shocks}},
  journal = {The Astrophysical Journal},
  volume = {288},
  pages = {342--352},
  year = {1985},
  doi = {10.1086/162797}
}

@article{Gary91,
  author = {{Gary}, S. P.},
  title = {{Electromagnetic ion/ion instabilities and their consequences in space plasmas: A review}},
  journal = {Space Science Reviews},
  volume = {56},
  pages = {373--415},
  year = {1991},
  doi = {10.1007/BF00196632}
}

@article{Skilling75,
  author = {{Skilling}, J.},
  title = {{Cosmic ray streaming -- I. Effect of Alfv\'{e}n waves on particles}},
  journal = {Monthly Notices of the Royal Astronomical Society},
  volume = {172},
  pages = {557--566},
  year = {1975},
  doi = {10.1093/mnras/172.3.557}
}

@article{Bell04,
  author = {{Bell}, A.~R.},
  title = {{Turbulent amplification of magnetic field and diffusive shock acceleration of cosmic rays}},
  journal = {Monthly Notices of the Royal Astronomical Society},
  volume = {353},
  pages = {550--558},
  year = {2004},
  doi = {10.1111/j.1365-2966.2004.08097.x}
}

@article{McKenzieVolk82,
  author = {{McKenzie}, J.~F. and {V{\"o}lk}, H.~J.},
  title = {{Non-linear theory of cosmic ray shocks including self-generated Alfv\'{e}n waves}},
  journal = {Astronomy and Astrophysics},
  volume = {116},
  pages = {191--200},
  year = {1982}
}

@article{McPherron72,
  author = {{McPherron}, R. L. and {Russell}, C. T. and {Coleman}, P. J.},
  title = {{Fluctuating magnetic fields in the magnetosphere. II: ULF waves}},
  journal = {Space Science Reviews},
  volume = {13},
  pages = {411--454},
  year = {1972},
  doi = {10.1007/BF00219165}
}

@article{Hollweg71,
  author = {{Hollweg}, J.~V.},
  title = {{Density fluctuations driven by Alfv\'{e}n waves}},
  journal = {Journal of Geophysical Research},
  volume = {76},
  pages = {5155--5161},
  year = {1971},
  doi = {10.1029/JA076i022p05155}
}

@article{Parker65,
  author = {{Parker}, E.~N.},
  title = {{The passage of energetic charged particles through interplanetary space}},
  journal = {Planetary and Space Science},
  volume = {13},
  pages = {9--49},
  year = {1965},
  doi = {10.1016/0032-0633(65)90131-5}
}

@book{Stix1992,
  author = {{Stix}, T.~H.},
  title = {{Waves in Plasmas}},
  publisher = {American Institute of Physics},
  address = {New York},
  year = {1992}
}

@article{Capanema26,
  author = {{Capanema}, A. and {Blasi}, P. and {Sobacchi}, E.},
  title = {{Acoustic instability at shock-wave precursors}},
  journal = {Astronomy and Astrophysics},
  volume = {710},
  pages = {A14},
  year = {2026},
  eprint = {2604.12514},
  archivePrefix = {arXiv}
}

@article{Whittlesey20,
  author = {{Whittlesey}, P.~L. and {Larson}, D.~E. and {Kasper}, J.~C. and {Halekas}, J. and {Abatcha}, M. and {Abiad}, R. and {Berthomier}, M. and {Case}, A.~W. and {Chen}, J. and {Curtis}, D.~W. and {Dalton}, G. and {Klein}, K.~G. and {Korreck}, K.~E. and {Livi}, R. and {Ludlam}, M. and {Marckwordt}, M. and {Rahmati}, A. and {Robinson}, M. and {Slagle}, A. and {Stevens}, M.~L. and {Tiu}, C. and {Verniero}, J.~L.},
  title = {{The Solar Probe ANalyzers---Electrons on the Parker Solar Probe}},
  journal = {The Astrophysical Journal Supplement Series},
  volume = {246},
  pages = {74},
  year = {2020},
  doi = {10.3847/1538-4365/ab7370}
}

@incollection{Schwartz98,
  author = {{Schwartz}, S.~J.},
  title = {{Shock and discontinuity normals, Mach numbers, and related parameters}},
  booktitle = {Analysis Methods for Multi-Spacecraft Data},
  editor = {{Paschmann}, G. and {Daly}, P.~W.},
  publisher = {ESA Publications Division},
  series = {ISSI Scientific Reports Series SR-001},
  volume = {1},
  pages = {249--270},
  year = {1998}
}

@article{CaprioliSpit14c,
  author = {{Caprioli}, D. and {Spitkovsky}, A.},
  title = {{Simulations of Ion Acceleration at Non-relativistic Shocks. III. Particle Diffusion}},
  journal = {The Astrophysical Journal},
  volume = {794},
  pages = {47},
  year = {2014},
  doi = {10.1088/0004-637X/794/1/47}
}

@article{TorrenceCompo98,
  author = {{Torrence}, C. and {Compo}, G.~P.},
  title = {{A Practical Guide to Wavelet Analysis}},
  journal = {Bulletin of the American Meteorological Society},
  volume = {79},
  number = {1},
  pages = {61--78},
  year = {1998},
  doi = {10.1175/1520-0477(1998)079<0061:APGTWA>2.0.CO;2}
}

@article{GarySmith09,
  author = {{Gary}, S.~P. and {Smith}, C.~W.},
  title = {{Short-wavelength turbulence in the solar wind: Linear theory of whistler and kinetic Alfv\'en fluctuations}},
  journal = {Journal of Geophysical Research},
  volume = {114},
  pages = {A12105},
  year = {2009},
  doi = {10.1029/2009JA014525}
}

@article{TenBarge12,
  author = {{TenBarge}, J.~M. and {Podesta}, J.~J. and {Klein}, K.~G. and {Howes}, G.~G.},
  title = {{Interpreting Magnetic Variance Anisotropy Measurements in the Solar Wind}},
  journal = {The Astrophysical Journal},
  volume = {753},
  pages = {107},
  year = {2012},
  doi = {10.1088/0004-637X/753/2/107}
}

@article{Howes12,
  author = {{Howes}, G.~G. and {Bale}, S.~D. and {Klein}, K.~G. and {Chen}, C.~H.~K. and {Salem}, C.~S. and {TenBarge}, J.~M.},
  title = {{The Slow-Mode Nature of Compressible Wave Power in Solar Wind Turbulence}},
  journal = {The Astrophysical Journal Letters},
  volume = {753},
  pages = {L19},
  year = {2012},
  doi = {10.1088/2041-8205/753/1/L19}
}

@article{Pedersen_1995,
  author       = {Pedersen, A.},
  title        = {Solar wind and magnetosphere plasma diagnostics by spacecraft electrostatic potential measurements},
  journal      = {Annales Geophysicae},
  volume       = {13},
  pages        = {118--129},
  year         = {1995},
  doi          = {10.1007/s00585-995-0118-8}
}

@article{Chen_2012a,
  title = {Density Fluctuation Spectrum of Solar Wind Turbulence between Ion and Electron Scales},
  author = {Chen, C. H. K. and Salem, C. S. and Bonnell, J. W. and Mozer, F. S. and Bale, S. D.},
  journal = {Phys. Rev. Lett.},
  volume = {109},
  issue = {3},
  pages = {035001},
  numpages = {6},
  year = {2012},
  month = {Jul},
  publisher = {American Physical Society},
  doi = {10.1103/PhysRevLett.109.035001}
}

@article{Mozer_2022,
doi = {10.3847/1538-4357/ac4f42},
year = {2022},
month = {feb},
publisher = {The American Astronomical Society},
volume = {926},
number = {2},
pages = {220},
author = {Mozer, F. S. and Bale, S. D. and Kellogg, P. J. and Larson, D. and Livi, R. and Romeo, O.},
title = {An Improved Technique for Measuring Plasma Density to High Frequencies on the Parker Solar Probe},
journal = {The Astrophysical Journal}
}

@article{Chen_2013b,
    author = {Chen, C. H. K. and Howes, G. G. and Bonnell, J. W. and Mozer, F. S. and Klein, K. G. and Bale, S. D.},
    title = {Kinetic scale density fluctuations in the solar wind},
    journal = {AIP Conference Proceedings},
    volume = {1539},
    number = {1},
    pages = {143-146},
    year = {2013},
    month = {06},
    issn = {0094-243X},
    doi = {10.1063/1.4811008}
}

@misc{Mondal_2026,
      title={The Nature of Turbulence at Sub-Electron Scales in the Solar Wind}, 
      author={Shiladittya Mondal and Christopher H. K. Chen and Davide Manzini},
      year={2026},
      eprint={2509.17061},
      archivePrefix={arXiv},
      primaryClass={physics.space-ph},
      url={https://arxiv.org/abs/2509.17061}, 
}

@inproceedings{DorfiDrury85,
	author = {{Dorfi}, E.~A. and {Drury}, L. O'C.},
	title = {{A cosmic ray driven instability}},
	booktitle = {International Cosmic Ray Conference},
	series = {International Cosmic Ray Conference},
	volume = {3},
	pages = {121},
	year = {1985}}

@article{Gedalinreview26,
  author  = {{Gedalin}, M. and {Balikhin}, M. and {Krasnoselskikh}, V.},
  title   = {{Collisionless Shocks: Contemporary State After Three Quarters of a Century of Research}},
  journal = {Reviews of Geophysics},
  volume  = {64},
  number  = {3},
  eid     = {e2025RG000893},
  pages   = {e2025RG000893},
  year    = {2026},
  doi     = {10.1029/2025RG000893}
}

@ARTICLE{Bamert04,
       author = {{Bamert}, K. and {Kallenbach}, R. and {Ness}, N.~F. and {Smith}, C.~W. and {Terasawa}, T. and {Hilchenbach}, M. and {Wimmer-Schweingruber}, R.~F. and {Klecker}, B.},
        title = "{Hydromagnetic Wave Excitation Upstream of an Interplanetary Traveling Shock}",
      journal = {\apjl},
         year = 2004,
       volume = {601},
        pages = {L99-L102},
          doi = {10.1086/381962}
}

@ARTICLE{LuttrellRichter87,
       author = {{Luttrell}, A. and {Richter}, A.~K.},
        title = "{A study of MHD fluctuations upstream and downstream of quasiparallel interplanetary shocks}",
      journal = {Journal of Geophysical Research},
         year = 1987,
       volume = {92},
        pages = {2243-2252},
          doi = {10.1029/JA092iA03p02243}
}

@ARTICLE{Kruparova25,
       author = {{Kruparova}, O. and {Szabo}, A. and {Jian}, L.~K. and others},
        title = "{Radial Evolution of Interplanetary Shock Properties with Heliospheric Distance: Observations from Parker Solar Probe}",
      journal = {\apjl},
         year = 2025,
       volume = {979},
        pages = {L10},
          doi = {10.3847/2041-8213/ada558}
}

@ARTICLE{Kennel84,
       author = {{Kennel}, C.~F. and {Scarf}, F.~L. and {Coroniti}, F.~V. and {Russell}, C.~T. and {Wenzel}, K.~P. and {Sanderson}, T.~R. and {van Nes}, P. and {Feldman}, W.~C. and {Parks}, G.~K. and {Smith}, E.~J. and {Tsurutani}, B.~T. and {Mozer}, F.~S. and {Temerin}, M. and {Anderson}, R.~R. and {Scudder}, J.~D. and {Scholer}, M.},
        title = "{Plasma and energetic particle structure upstream of a quasi-parallel interplanetary shock}",
      journal = {Journal of Geophysical Research},
         year = 1984,
       volume = {89},
        pages = {5419-5435},
          doi = {10.1029/JA089iA07p05419}
}

@article{Wilson09,
	author = {{Wilson}, L.~B., III and {Cattell}, C.~A. and {Kellogg}, P.~J. and {Goetz}, K. and {Kersten}, K. and {Kasper}, J.~C. and {Szabo}, A. and {Meziane}, K.},
	title = {{Low-frequency whistler waves and shocklets observed at quasi-perpendicular interplanetary shocks}},
	journal = {Journal of Geophysical Research},
	volume = {114},
	pages = {A10106},
	year = {2009},
	doi = {10.1029/2009JA014376}
}

@article{Wilson12,
	author = {{Wilson}, L.~B., III and {Koval}, A. and {Szabo}, A. and {Breneman}, A. and {Cattell}, C.~A. and {Goetz}, K. and {Kellogg}, P.~J. and {Kersten}, K. and {Kasper}, J.~C. and {Maruca}, B.~A. and {Pulupa}, M.},
	title = {{Observations of electromagnetic whistler precursors at supercritical interplanetary shocks}},
	journal = {Geophysical Research Letters},
	volume = {39},
	pages = {L08109},
	year = {2012},
	doi = {10.1029/2012GL051581}
}

@article{Oka19,
	author = {{Oka}, M. and {Otsuka}, F. and {Matsukiyo}, S. and {Wilson}, L.~B., III and {Argall}, M.~R. and {Amano}, T. and {Phan}, T.~D. and {Hoshino}, M. and {Le Contel}, O. and {Gershman}, D.~J. and {Burch}, J.~L. and {Torbert}, R.~B. and {Dorelli}, J.~C. and {Giles}, B.~L. and {Ergun}, R.~E. and {Russell}, C.~T.},
	title = {{Electron Scattering by Low-frequency Whistler Waves at Earth's Bow Shock}},
	journal = {\apj},
	volume = {886},
	pages = {53},
	year = {2019},
	doi = {10.3847/1538-4357/ab4a81}
}

\end{document}